\documentclass[12pt]{article}
\usepackage{graphicx} 
\usepackage{subcaption}
\usepackage{setspace}
\usepackage{epstopdf}
\usepackage{tikz}
\usetikzlibrary{arrows,automata}
\usetikzlibrary{patterns}
\usepackage{latexsym,amssymb,amsmath,amsthm,amsfonts}
\usepackage{comment}
\usepackage{url}
\usepackage{dsfont}
\usepackage{ragged2e} 
\usepackage{multirow}
\usepackage{listings}
\usepackage{xcolor}
\usepackage{titlesec}
\usepackage{pgfplots}
\usepackage{authblk}
\usepackage{dsfont}
\usepackage{anyfontsize} %
\usepackage{colortbl} 
\usepackage{xcolor} 
\usepackage{steinmetz}
\usepackage{graphicx} 
\usepackage{caption}  
\usepackage{subcaption} 
\usepackage{graphicx}
\usepackage{hyperref}
\usepackage{enumitem}

\usepackage{pgfkeys}

\usepackage{parskip}

\usepackage{appendix}
\usepackage{tikz}
\usetikzlibrary{automata, positioning} 
\usepackage{float} 
\usepackage[round]{natbib}
\usepackage{titlesec}
\titleformat{\section}{\normalfont\Large\bfseries}{\thesection}{1em}{}
\usepackage{tocloft}

\usepackage{tabularx}
\usepackage{parskip}

\usepackage{hyperref}
\hypersetup{
    colorlinks=true,
    linkcolor=blue,
    filecolor=magenta,      
    urlcolor=blue, 
    pdftitle={Overleaf Example},
    pdfpagemode=FullScreen,
}

\usepackage[a4paper, total={6in, 8in}]{geometry}

\usepackage{booktabs}
\usepackage{multirow}
\usepackage{caption}
\usepackage{tabularx}
\usepackage{amsmath}
\usepackage{cite}
\usepackage{mathptmx} 

\usepackage{fancyhdr}
\usepackage{graphicx} 
\usepackage{subcaption}
\usepackage{setspace}
\usepackage{epstopdf}
\usepackage{tikz}
\usetikzlibrary{arrows,automata}
\usetikzlibrary{patterns}
\usepackage{latexsym,amssymb,amsmath,amsthm,amsfonts}
\usepackage{comment}
\usepackage{url}
\usepackage{dsfont}
\usepackage{ragged2e} 
\usepackage{multirow}
\usepackage{listings}
\usepackage{xcolor}
\usepackage{titlesec}

\usepackage{graphicx} 
 \usepackage{amsmath}

\title{
{Bayesian measurement error modeling of latent time series structure to assess the impact of pollutants on health}\\~\\}
\author[1]{Yanfei Qu}
\author[1]{David A. Stephens}
\affil[1]{Department of Mathematics and Statistics, McGill University}

\date{}

\begin{document}

\clearpage\maketitle
\thispagestyle{empty}

\addcontentsline{toc}{section}{Abstract}
\section*{Abstract}
The association between levels of air pollution and mortality rate is well-established, but quantifying the magnitude of the effect is sometimes complicated by limitations in the data. In this paper, a joint Bayesian hierarchical model is developed to identify significant predictor factors and quantify their impacts on mortalities. To account for potential measurement error in the pollution data, the observed pollutant levels were treated as noisy proxies for the true exposure, therefore requiring a measurement error structure.  We illustrate the developed model by performing an analysis of the association between weekly air pollution levels and cardiovascular and respiratory mortality in Los Angeles (LA) County over five years from January 2018 to December 2022. The purpose of the study was to quantify the impact of the main air pollutants (PM\textsubscript{2.5}, PM\textsubscript{10}, SO\textsubscript{2}, NO\textsubscript{2}, CO, and O\textsubscript{3}) and of temperature on weekly mortality from four causes: chronic obstructive pulmonary diseases, pneumonia, heart failures, and malignant neoplasms.  The results, supported by Bayesian model comparison criteria, indicated that weekly county-level cause-specific mortalities were significantly associated with certain ranges of air pollutants, and that some of these cause-specific mortalities
had significant associations with ambient temperature levels. The analysis revealed that several pollutants appeared to be associated with lower mortality; we interpreted this counterintuitive finding from various perspectives.

\textit{keywords}: Bayesian Hierarchical Model, State Space Model, Measurement Error, Time series
\section{Introduction}\label{sec1}
As air pollution became an increasingly significant global public health concern, numerous studies have documented the adverse effects of air pollutants on human health. 
Despite this wide agreement, data on which these conclusions are based are typically messy, incomplete, and complex; pollution monitoring networks are unevenly distributed, some pollutants are measured sparsely or intermittently, and health effects are typically not due to exposure to a single pollutant, but are the result of the joint impact of multiple pollutants. In this paper, we build a joint Bayesian hierarchical model that reflects key aspects of this complexity. The model works in two stages: first, it treats observed pollutant levels as noisy proxies for true exposure rather than perfect measures, through a measurement error model; secondly, the model links those exposure estimates to cause-specific mortality outcomes across time. This structure allows us to account for measurement error directly and also borrow strength across time periods. We analyze time series of regional aggregate mortality counts and attempt to assess the link between true aggregate exposure and these health outcomes.

Widespread efforts have been made to quantify the effects of different types of pollutants on cause-specific mortality and hospitalizations, with particular emphasis on cardiovascular and respiratory conditions. The pollutants that have been most commonly studied included fine particulate matter (PM\textsubscript{2.5}), coarse particulate matter (PM\textsubscript{10}), sulphur dioxide (SO\textsubscript{2}), nitrogen dioxide (NO\textsubscript{2}), carbon monoxide (CO), and ozone (O\textsubscript{3}). These pollutants are routinely monitored by relevant governmental agencies worldwide, and the data are made publicly available. Among the pollutants, PM\textsubscript{2.5} has been found, due to its small aerodynamic diameter, to have the ability to penetrate deeply into the respiratory tract and enter the circulatory system, triggering systemic inflammatory responses and other pathophysiological processes \citep{xing2016impact}. Epidemiological and statistical studies, including the Harvard Six Cities Study   \citep{dockery1993association}, the large national time-series analyses of Medicare beneficiaries in the United States \citep{dominici2006fine}, and the 20 U.S. cities study \citep{samet2000fine}, have showed that exposure to PM\textsubscript{2.5} was associated with increased risks of all health outcomes except injuries. Other pollutants have also been found to have the potential to compound the health burdens by triggering inflammatory responses and oxidative stress, associating with increased risk for adverse cardiopulmonary events   \citep{samet2000fine, chen2007outdoor}. One study \citep{bell2004ozone} showed that a 10-ppb increase in the previous week's ozone was associated with a 0.52\% increase in daily mortality and a 0.64\% increase in cardiovascular and respiratory mortality in 95 US urban communities. Through hierarchical models implemented in two stages, Samoli \citep{samoli2007short} found a 1-$mg/m^3$ increase in the 2-day mean of CO levels was associated with a 1.20\% increase in total deaths and a 1.25\% increase in cardiovascular deaths. Zhang   \citep{zhang2011study} demonstrated that for urban air pollution, using a single-pollutant, current-day analysis, the relative risks (RR) for a 10 $\mu g/m^3$ increase in pollutant levels were: for respiratory disease mortality, 1.001 for PM\textsubscript{10}, 1.00003 for SO\textsubscript{2}, and 1.009 for NO\textsubscript{2}; for cardiovascular disease mortality, 1.002 for PM\textsubscript{10}, 1.000 for SO\textsubscript{2}, and 1.003 for NO\textsubscript{2} among citizens in Chaoyang District of Beijing. Lee \citep{lee2015carbon} conducted a population-based, longitudinal cohort study in Taiwan and concluded that patients with CO poisoning was associated with a higher risk of developing subsequent cardiovascular diseases (CVDs).
Zhang      \citep{zhang2017association} showed that in Hefei, China, for every increase of 10 $\mu g/m^3$ in SO\textsubscript{2}, NO\textsubscript{2}, and PM\textsubscript{10} levels, cardiovascular disease mortality increased by 5.26\%, 2.71\%, and 0.68\%, respectively, at lag03 (the cumulative moving average of a pollutant over the current day and the three preceding days). Recently Zhou \citep{Zhou2023} reported that in Ningbo, China, 
for every 10-unit increase in lag03 air pollutant concentration, hospitalization for pneumonia and asthma due to PM\textsubscript{2.5}, PM\textsubscript{10}, SO\textsubscript{2}, and NO\textsubscript{2} increased by 2.22\%, 1.94\%, 11.21\% and 5.42\%, respectively.

Generalized linear models (GLMs) have been widely used for estimating the association between pollution and health \citep{liu2026air, tango1994effect, he2006comparative}. Generalized additive models (GAMs) offer a more flexible alternative, allowing for nonlinear regression and accommodating exposure–response relationships \citep{he2006comparative, dominici2002use, hastie1987generalized}.
Dynamic Linear Models (DLMs), along with their Bayesian extensions, offer a flexible framework for air pollution epidemiology allowing for adaptation over time,  dynamically handling shifting trends, varying pollution effects, and temporal dependencies that show up in the measurements
\citep{lee2008modelling}.
These approaches span a spectrum from conventional regression to more flexible and dynamic specifications, each suited to different aspects of air pollution health research. Most studies have focused on either very short-term (e.g., daily) or long-term (e.g., annual) exposure windows \citep{dockery1993association,chen2025impact}, leaving intermediate time scales underexplored. Daily data offer fine temporal resolution, yet they are susceptible to high-frequency noise from transient influences such as rapid weather fluctuations or episodic events like wildfires. By comparison, aggregated summaries, such as the annual averages, tend to smooth away too much of the variability, thereby removing key seasonal structures in the data. Several papers focus on intermediate-frequency inputs,  such as the monthly averages used in Portuguese metro regions   \citep{duarte2025impact}, but work using weekly data is still rarely seen. Moreover, earlier daily-frequency investigations, including the 20 U.S. Cities Study   \citep{samet2000fine}, often reported that health impacts show up after about a 2- to 5-day delay. Motivated by this, we assessed the health consequences of air pollution using a one‑week time unit; weekly aggregation helps dampen measurement noise while still keeping the timing patterns that can get diluted when moving to monthly or yearly reporting. Thus, a weekly window represents a reasonable compromise for studying delayed health responses.

An important issue that arises during the study of air pollution problems is that of measurement error present in unadjusted exposure data \citep{sheppard2012confounding, dominici2000measurement}. Although monitoring systems can measure pollution with great precision at certain fixed locations, this measurement can be considered as an imprecise measure of true aggregate exposure at the population level.  When mortality data are aggregated to a region-wise summary, measurements at individual stations must be combined to inform a region-wise measure of pollution.  The framework of measurement error modeling can be used, and the unobserved true effective (aggregate) pollution level can be inferred from the observed pollution data. While existing approaches to fix measurement error in air pollution epidemiology have mostly treated exposure error as static or spatially-driven, they have largely overlooked the temporal correlation structure inherent in exposure measurements. The novelty in our work is that we simultaneously modeled both the measurement error and its serial dependence over time, so we acknowledged that errors on nearby days tend to be correlated, and then we developed correction methods that explicitly incorporate this temporal dimension.

The remainder of this paper is organized as follows. Section~\ref{Data} describes the study region and the data sources via which we illustrate the importance of time-indexed latent measurement error modeling. Section~\ref{Method} outlines the modeling framework. Section~\ref{simulation} presents simulation studies designed to evaluate the performance of the proposed models and to assess the robustness of the estimates under various data-generating scenarios. Section~\ref{Real Data Analysis} provides a descriptive analysis of the real data and applies the proposed model to the real data;
specifically, section~\ref{Results} reports the main empirical findings. Finally, Section~\ref{Discussion} interprets the results, discusses their implications, and suggests directions for future research.

\section{Study Area and Dataset}\label{Data}
\subsection{Study Area: Los Angeles County, California, United States}
\begin{figure}[t]
  \centering
  \includegraphics[width=0.75\linewidth]{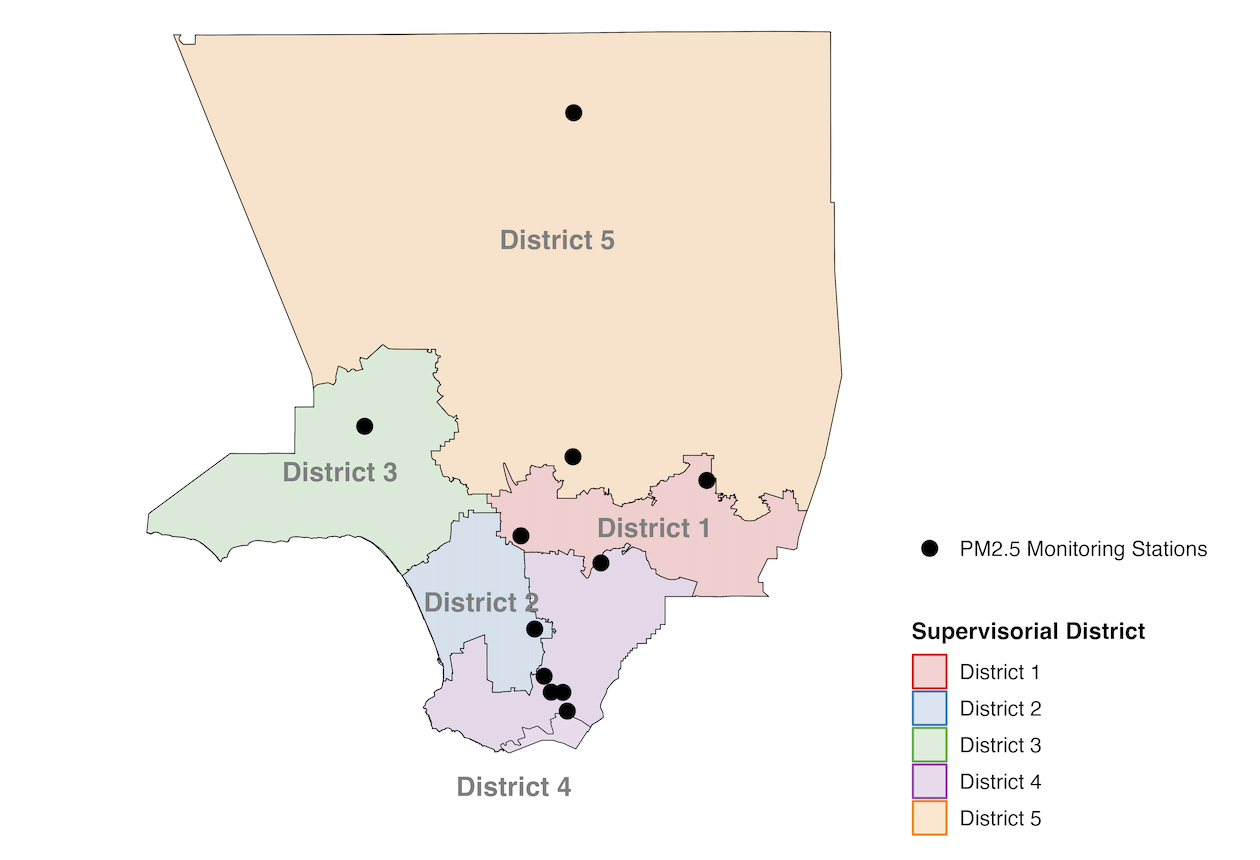}
  \caption{Los Angeles County Map with locations of monitoring stations (Santa Catalina Island and San Clemente Island are omitted from this map.)}
  \label{fig:Map}
\end{figure}
Los Angeles County provides a suitable setting for a study of the impact of pollution on aggregate mortality levels that incorporates latent measurement error modeling. Located in California, in the western United States, the LA county population declined from nearly 10 million in early 2018 to approximately 9 million by late 2022; the annual population of our studied time period is summarized in Table \ref{tab:Population}; county-level annual population data were sourced from the \href{https://fred.stlouisfed.org/}{Federal Reserve Economic Data (FRED)}.

\begin{table}[h]
\centering
\caption{Annual Population of Los Angeles County}
\label{tab:Population}
\begin{tabular}{c r}
\toprule
\textbf{Year} & \multicolumn{1}{c}{\textbf{Population (Thousands)}} \\
\midrule
2018 & 10061.533\\
2019 & 10011.602\\
2020 & 9996.634\\
2021 & 9809.239\\
2022 & 9748.447\\
\bottomrule
\end{tabular}
\end{table}

It has long been known for having some of the highest pollution levels in the U.S., mostly due to heavy traffic and geography that keeps exhaust and other emissions close to the ground, in part through temperature inversions  \citep{robinson1952some}. These conditions create ongoing threats to public health.   In LA county, the population is heavily concentrated in the southern and southwestern areas. Figure~\ref{fig:Map} represents the locations of monitoring stations that reported PM\textsubscript{2.5} measurements; most are situated in the southern part of the county. The county is divided into five supervisorial districts, each representing roughly two million constituents, also with the majority of these districts located in the county's southern region. Therefore, although the monitoring stations are concentrated in the south and sparse in the north, when aggregating for our real data analysis, this imbalance was mitigated by the fact that population density is also much higher in the south. Consequently, the averaged pollution level, while weighted toward southern conditions, reasonably reflects the exposure of the majority of residents. Given that our outcome was the total county-level cause-specific death counts, this aggregation should not introduce considerable bias.

\subsection{Dataset} 

\begin{figure}[tp]
    \setkeys{Gin}{width=1\linewidth,height=0.15\textheight}
    \centering
    \begin{minipage}{0.45\linewidth}
        \centering
        \subfloat[PM\textsubscript{2.5}]{\includegraphics{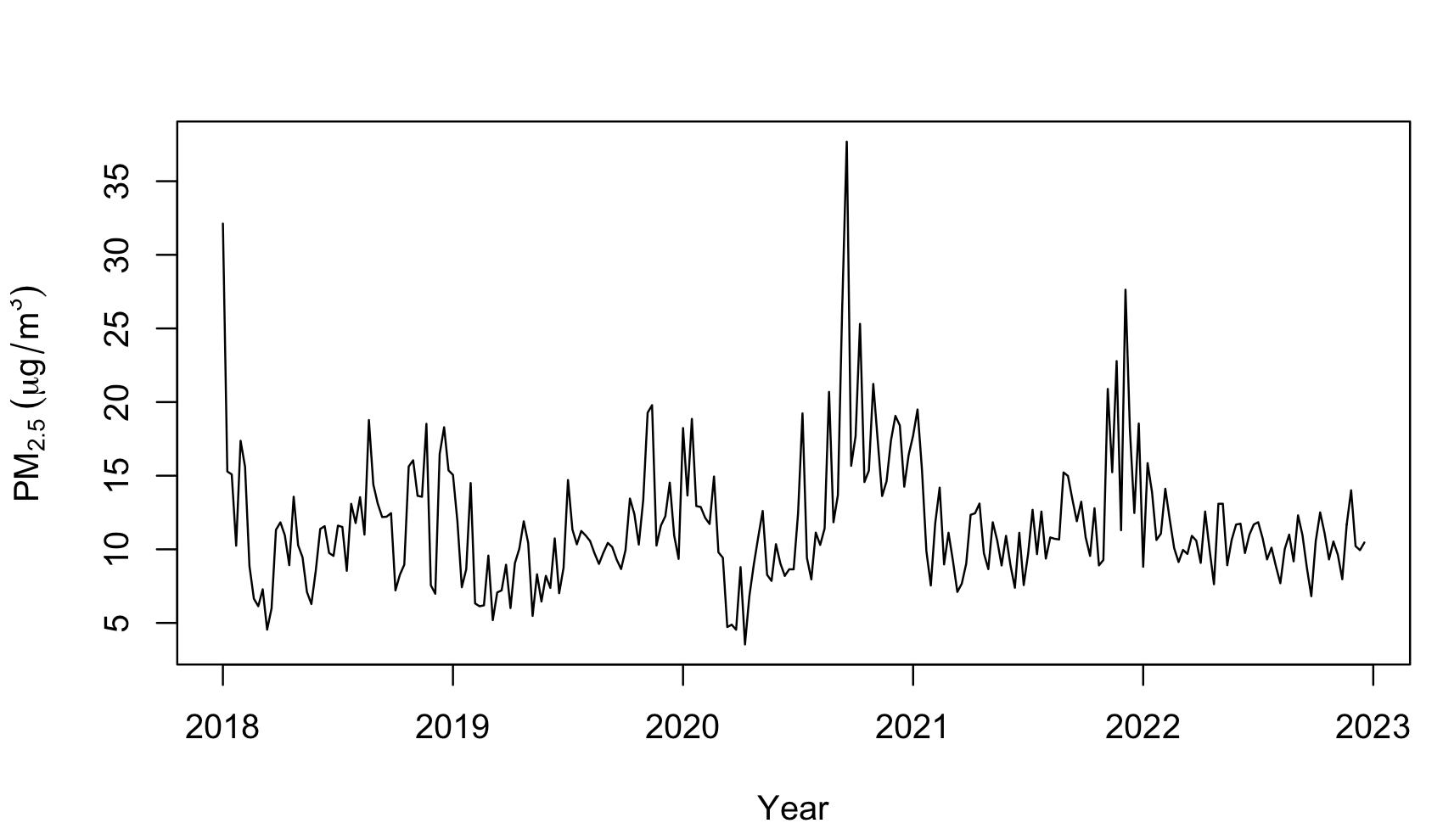}\label{fig:PM2.5}}
    \end{minipage}
    \quad
    \begin{minipage}{0.45\linewidth}
        \centering
        \subfloat[PM\textsubscript{10}]{\includegraphics{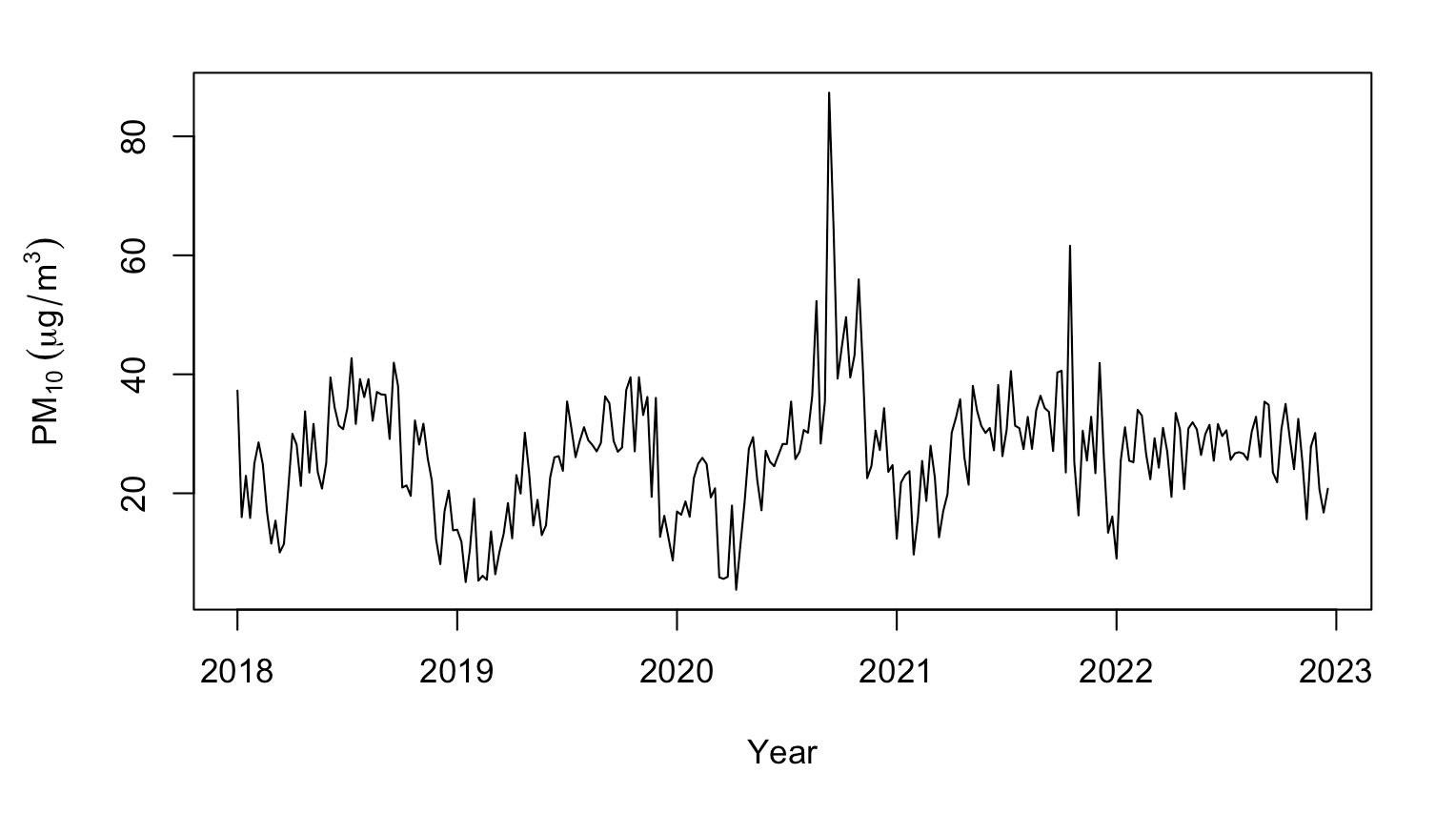}\label{fig:PM10}}
    \end{minipage}
    
    \vspace{1ex}
    
    \begin{minipage}{0.45\linewidth}
        \centering
        \subfloat[SO\textsubscript{2}]{\includegraphics{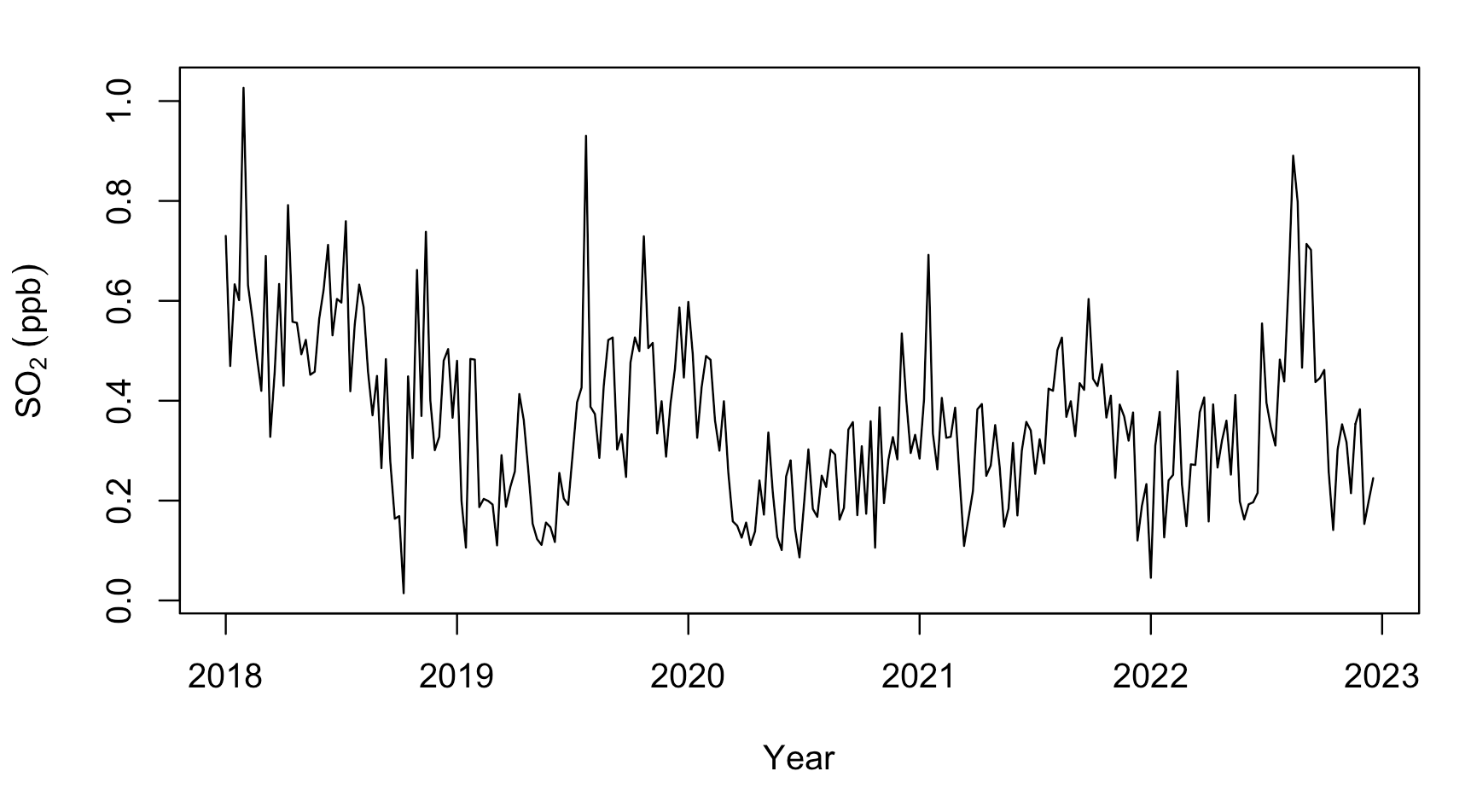}\label{fig:SO2}}
    \end{minipage}
    \quad
    \begin{minipage}{0.45\linewidth}
        \centering
        \subfloat[NO\textsubscript{2}]{\includegraphics{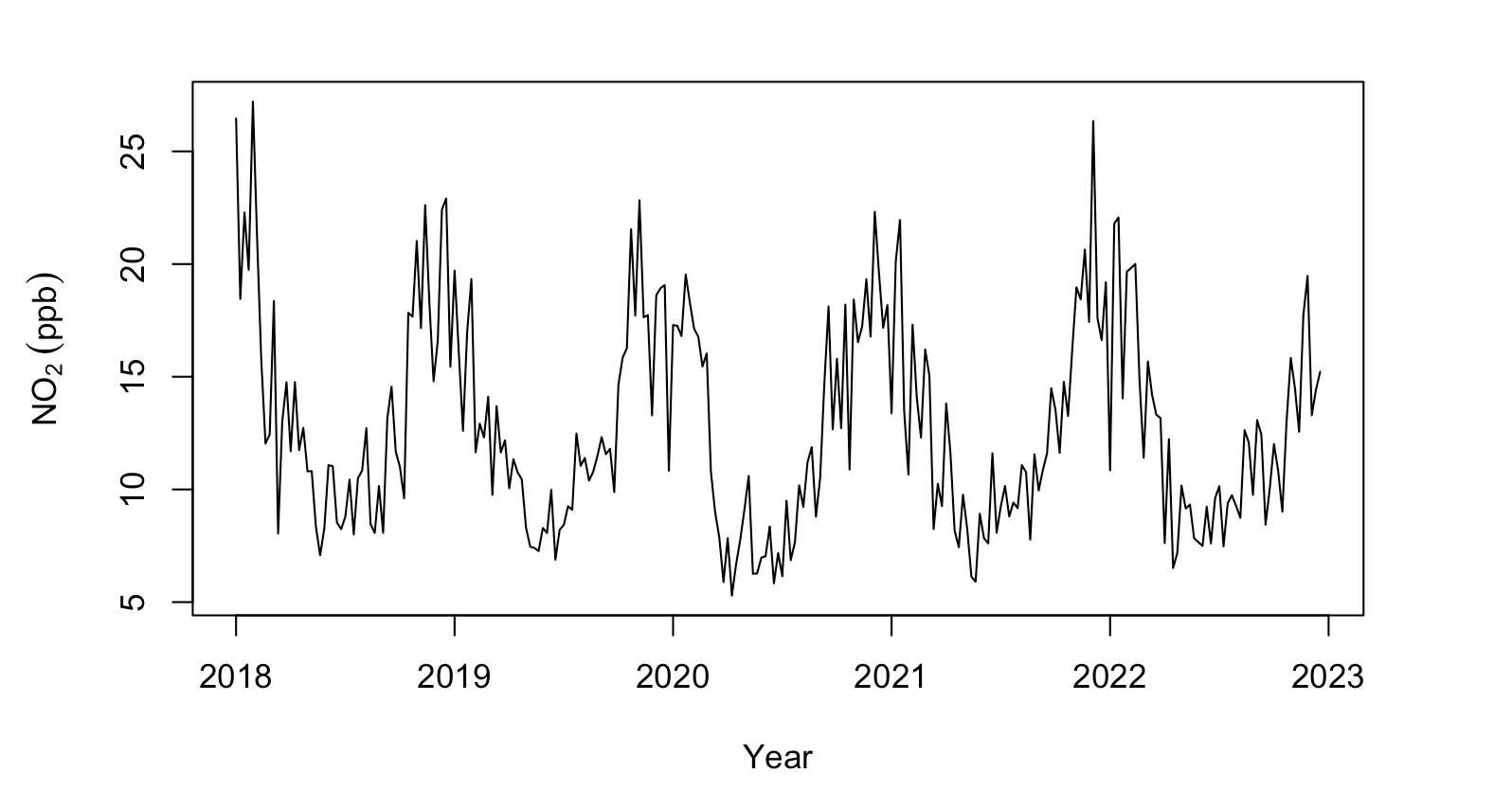}\label{fig:NO2}}
    \end{minipage}
    
    \vspace{1ex}
    
    \begin{minipage}{0.45\linewidth}
        \centering
        \subfloat[CO]{\includegraphics{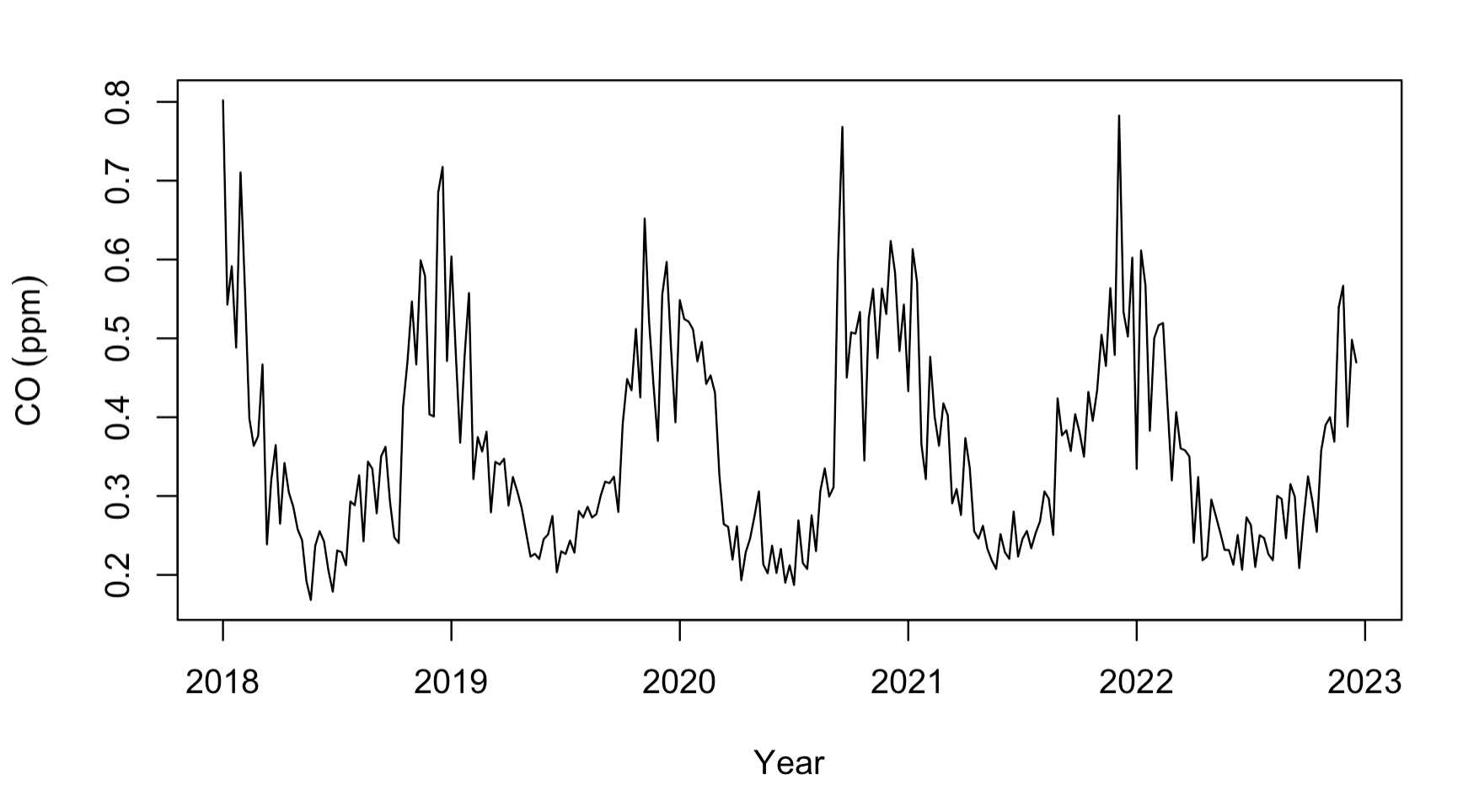}\label{fig:CO}}
    \end{minipage}
    \quad
    \begin{minipage}{0.45\linewidth}
        \centering
        \subfloat[O\textsubscript{3}]{\includegraphics{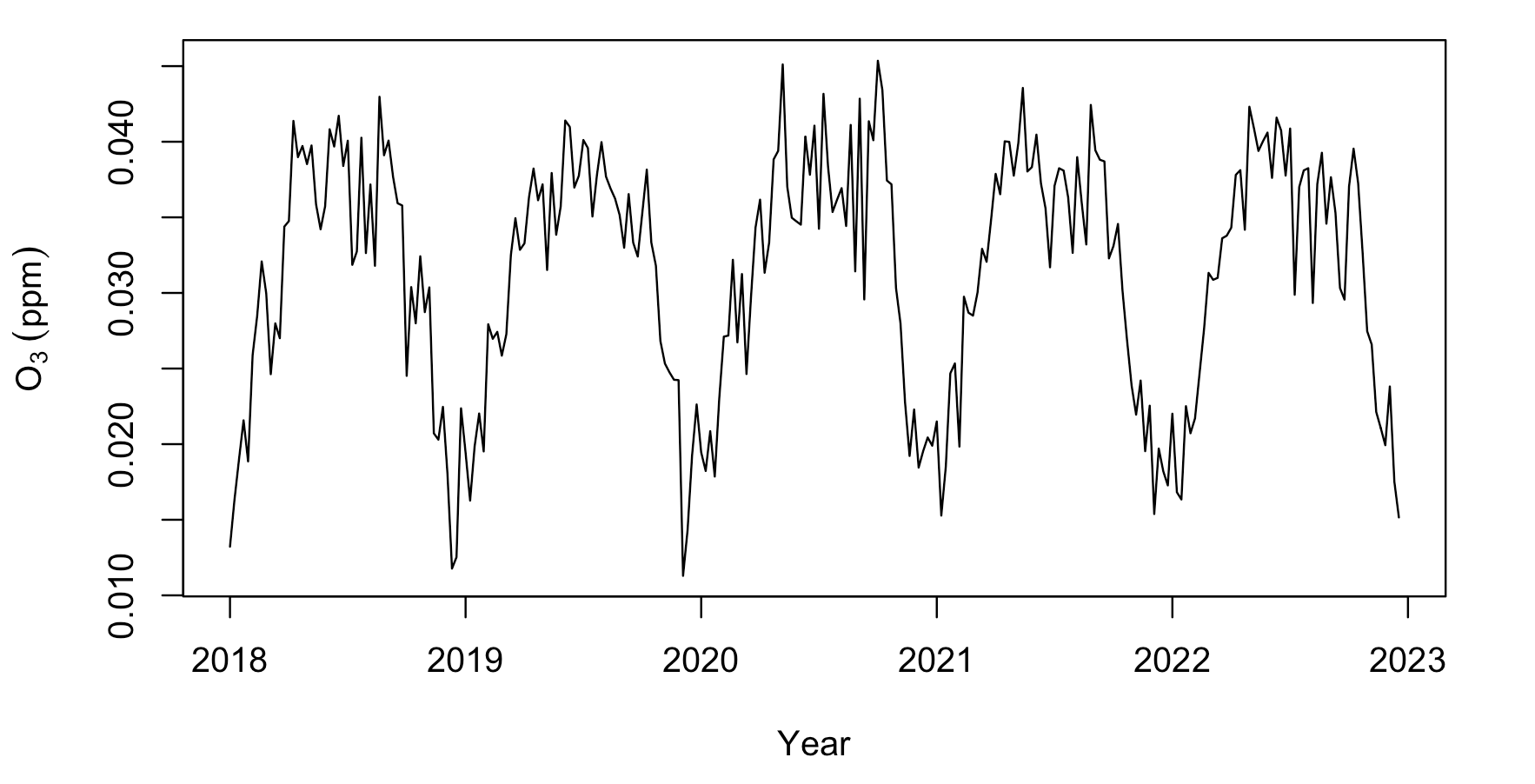}\label{fig:O3}}
    \end{minipage}
    
    \caption{Time series of measured air pollutants in LA County (January 1, 2018 - December 17, 2022).  Average measures across monitoring stations.}
    \label{fig:Air pollutants plots}
\end{figure}

Air quality monitoring stations, accredited by the \href{https://aqs.epa.gov/aqsweb/airdata/download_files.html}{U.S. Environmental Protection Agency (EPA)}, have been established throughout LA county to collect daily measurements of air pollution data. For our analysis, we aggregated the data from all available stations to construct consistent time series aligned with the Morbidity and Mortality Weekly Report (MMWR) weeks specific to this study\footnote{MMWR weeks are defined as the 
standardized epidemiologic week used by the Centers for Disease Control and Prevention (CDC) and public health departments. An MMWR week always spans from Sunday through Saturday. Weeks are numbered from week 1 to week 52 or 53.}. To match this weekly temporal resolution, we calculated the 7-day average of all valid daily measures within each week. The weekly pollution time series (averaged across monitoring stations)  are shown in Figure \ref{fig:Air pollutants plots}, and comprise $N=259$ observations.

\label{dataset}

Daily mean temperatures for Los Angeles County were obtained from the \href{https://www.ncei.noaa.gov/data/nclimgrid-daily/access/averages/}{NOAA NClimGrid Daily Averages} dataset. Following the same approach used for the air pollution data, we calculated weekly mean temperatures by averaging daily values over each MMWR week.

MMWR Weekly population estimates were estimated from annual measures using simple linear interpolation. Further details on this conversion are provided in \textbf{Supplementary Material} Section 1. The health outcome data, in our case, the MMWR weekly county-level counts of COPD, pneumonia, heart failure, and neoplasms deaths, were accessed via the \href{https://wonder.cdc.gov/mcd-icd10-provisional.html}{CDC WONDER's Provisional Mortality Statistics, 2018–Last Week}  from January 1, 2018, to December 17, 2022. 

\section{Analysis Methodology} \label{Method}

We examined how a one-week lag in air pollution levels, together with the current temperature, might shape health outcomes. We chose this one-week window because health effects from exposure are unlikely to be immediate; rather, they tend to emerge after several days, and we would like to capture that delayed response. We developed a Bayesian hierarchical modeling framework. At the first level, we modeled health outcomes using a Poisson log‑linear regression; this was a natural choice for count data such as hospital visits or mortalities, but could easily be extended to other count data models. At the second level, we modeled pollution measurements using a state‑space model under the assumption that direct monitor readings cannot perfectly reflect population‑level exposure. By integrating these two components, we aimed to obtain a credible estimate of the joint effects of temperature and pollution on health in this complex urban environment.

\subsection{Intuitive Model Hierarchy}

In reality, ambient air pollutant concentrations do not vary in isolation but tend to fluctuate simultaneously due to shared sources
found to be closely linked to one another over a certain period
  \citep{zhu2024review}. This natural co-variation means that pollution levels across different species are inherently correlated over time. A Vector Autoregressive (VAR) model explicitly captures these cross-pollutant dependencies, allowing the past levels of all pollutants to inform the current state of each pollutant, which seemed to be a realistic structure. We let $Z_t = (Z_{1t}, \ldots, Z_{Kt})'$ denote the true (unobserved) log-transformed and centered pollution concentrations for $K$ pollutants at time $t \in 1, \ldots, N$. The observed pollution data $Y_t$ is subject to measurement error:
\begin{equation*}
Y_t = Z_t + \boldsymbol{\varepsilon}_t, \quad \boldsymbol{\varepsilon}_t \sim \text{MVN}(0, \Sigma_\varepsilon)
\end{equation*}
where $\Sigma_\varepsilon = \text{diag}(\sigma_{\varepsilon,1}^2, \ldots, \sigma_{\varepsilon,K}^2)$ is a diagonal measurement error covariance matrix. The latent pollution process evolves according to a full VAR(1) model:
\begin{equation*}
Z_t = \mu_t + \Phi(Z_{t-1} - \mu_{t-1}) + \eta_t, \quad \eta_t \sim \text{MVN}(0, \Sigma_\eta)
\end{equation*}
where $\Phi$ is a $K \times K$ matrix of autoregressive coefficients capturing both own-lag and cross-lag effects, and $\Sigma_\eta$ is the $K \times K$ process error covariance matrix allowing correlations across pollutants. The mean structure $\mu_t = a + b  t + S_t$ captures deterministic components including intercepts $a = (a_1, \ldots, a_K)'$, linear trends $b = (b_1, \ldots, b_K)'$, and seasonal harmonics:

\begin{equation*}
S_t = 
\beta^{\text{sin}} \sin\left(\frac{2\pi  t}{52}\right) + \beta^{\text{cos}} \cos\left(\frac{2\pi  t}{52}\right) 
\end{equation*}

The health outcome model links the true pollution levels to cause-specific mortality, which follows the Poisson ($\lambda_t$) distribution with
\begin{equation*}
\log(\lambda_t) = \alpha + \beta Z_{t-1} + \gamma  \text{T}_{t-1} + \log(O_t)
\end{equation*}
where $T_t$ is the recorded average temperature in week $t$, $\beta = (\beta_1, \ldots, \beta_K)$ are the pollution effect coefficients, $\alpha$ is an intercept, $\gamma$ captures the temperature effect, and $\log(O_t)$ serves as an offset, in our case, the population count.

The model was implemented using MCMC with 3 parallel chains, 100,000 burn-in iterations, and 3,000 posterior samples. Even after extensive tuning, the MCMC algorithm showed persistent convergence difficulties due to the complex posterior form, and the weak identifiability of the latent states and the VAR parameters. With the identifiability issues in mind, we examined alternative approaches and models aimed at simplification and dimension reduction.

\subsection{Singular Value Decomposition (SVD) Transform of the Observed Pollution Data}\label{SVD Transformation}
The full VAR(1) model allows for cross-sectional and temporal dependence between the latent series, and the inferred true value series then influences the outcome via the Poisson model.  However, this latent series has a complex structure, and so to simplify this, we employed a Singular Value Decomposition (SVD) transformation on the log-transformed and centred pollution data (\textbf{Supplementary Materials} Section 2).
The SVD decomposition assumes that the correlation structure among the $K$ pollution series is constant over time, and reduces the pollution data to essentially uncorrelated series. That is, while each individual series exhibits temporal autocorrelation (which is subsequently modeled through the AR(1) structure, described in Section \ref{state space model}), the contemporaneous relationships between different pollutants are assumed stable across the study period.  In the pre-processing, the initial log-transformation and centring were performed to normalize the distribution and improve the interpretability of the intercepts, while the subsequent SVD transformation was used to reduce multicollinearity. All dimensions were retained throughout the process to preserve the full variation present in the original data. After the MCMC is implemented on the SVD-transformed scale, the sampled values are back-transformed onto the original scale for entry into the outcome model (see Section \ref{Approach 1}).


We let $Y\in \mathbb{R}^{N \times K}$ represent the log-transformed and centred observed pollution levels. At this step, we include all six pollutants ($K=6$), indexed in the order: PM$_ {2.5}$ (1), PM$_ {10}$ (2), SO$_ {2}$ (3), NO$_ {2}$ (4), CO (5), and O$_ {3}$ (6). 
The SVD transformation of $Y$ is given by $Y = U \, \Sigma \, V^\top$, where $U \in \mathbb{R}^{N \times N}$ is the matrix of left singular vectors associated with the temporal dimension, $\Sigma \in \mathbb{R}^{N \times K}$ is a diagonal matrix containing the singular values $\sigma_1 \ge ... \ge \sigma_K \ge 0$, and $V \in \mathbb{R}^{K \times K}$ is the matrix of right singular vectors associated with the pollutant variables. The objective of this decomposition was to rotate the original pollutant coordinate system into a new set of axes that are orthogonal.
The transformation matrix denoted $Q \equiv V$, which acts on the columns of $Y$ and thus transforms the pollutant variables rather than the time index. The transformed pollution matrix was then $Y^* = Y Q$
where $Y^*\in \mathbb{R}^{n \times K
}$ represents the pollution level in the rotated latent space. Equivalently, the transformed value of pollutant $k$-th ($k\in\{1,...,K\}$) at week $t$ could be written as
$Y^*_{t,k} = \sum_{j=1}^K Y_{t,j} Q_{j, k }$
where 
$Y_{t,j}$ denotes the log-transformed and centred pollution level of pollutant $j\in\{1,...,K\}$ at week $t\in\{1,...,N\}$, $Q_{j,k}$ represents the element at the $j$-th row and $k$-th column of the transformation matrix Q, and 
$Y^*_{t,k}$ is the SVD-transformed observed pollution level of pollutant $k$ at week $t$. 

\subsection{State-Space Modeling in the Transformed Space} \label{state space model}


Let $Z^{\ast}_{t,k}$ denote the true latent pollution level 
in the SVD-transformed space, 
with $k$ being the pollutant index and $t$ being the time point index. Each transformed pollution component, obtained via SVD, largely removed the cross-sectional dependence across pollutants. To capture the potential temporal dynamics of the pollution levels, the latent pollution process $\{ Z^{\ast}_{k}\}$ was modeled independently for each $k$ using a first-order autoregressive process, AR(1). To capture seasonal and periodic patterns in pollutant levels, we included sine and cosine functions as part of the time-varying mean. These functions allowed the model to flexibly represent cycles of any phase and frequency, while remaining mathematically convenient due to their orthogonality. Specifically, the harmonic terms accounted for any predictable, repeating variation, while the state-space model simultaneously handled temporal dependence and measurement error.  We set the frequency of the sinusoidal terms to $1/52$, reflecting our assumption that the underlying cycle repeats on a 52-week basis.
\begin{equation*}
    \mu_{t,k}=a_k + b_k  t + \beta_{1,k} \sin\left(\frac{2\pi}{52}t\right) + \beta_{2,k} \cos\left(\frac{2\pi}{52}t\right)
\end{equation*}
where $\mu_{t,k}$ is the time-varying mean of the latent true level of pollutant $k$ at time $t$, $a_k$ is the pollutant-specific intercept capturing the baseline level of component $k$, $b_k$ is the pollutant-specific linear temporal trend, sin$(2\pi t/52)$ and cos$(2\pi t/52)$ are the predefined harmonic sinusoidal seasonal basis functions evaluated at time $t$, $\beta_{1k}$ and $\beta_{2k}$ are the corresponding seasonal coefficients. Let $\tau_{\text{proc},k}$ be the process precision (inverse variance). For $t=1$,  $Z^{\ast}_{1,k} \sim \mathcal{N}\!\Big(\mu_{1,k}, \tau_{\text{proc},k}^{-1} \Big)$.
For subsequent time points $t\in\{2,\dots,N\}$, the state evolved according to
where $\phi_k$ is the autoregressive parameter governing temporal dependence in component $k$:
\begin{equation*}
Z^{\ast}_{t,k} \sim \mathcal{N}\!\Big(\mu_{t,k}+\phi_{k}(Z^*_{t-1,k}-\mu_{t-1,k}),\, \tau_{\text{proc},k}^{-1} \Big)
\end{equation*}


As defined in Section \ref{SVD Transformation}, let the SVD-transformed pollution level $Y^{\ast}_{t,k}$ be the noisy measurement of the latent true level $Z^{\ast}_{t,k}$, reflecting the fact that in reality, the true pollution exposure differs from what's recorded by the monitoring stations. 
\begin{equation}\label{Measurement Error Model}
Y^{\ast}_{t,k} = Z^{\ast}_{t,k} + \varepsilon_{t,k}, \quad \varepsilon_{t,k} \sim \mathcal{N}\!\Big(0, \tau_{\text{me},k} ^{-1}\Big),
\end{equation}
where 
$\tau_{\text{me},k}=1/\sigma^2_{me, k}$ is the measurement error precision for component $k$ when $\sigma^2_{me,k}$ denotes the variance of the measurement error.  Equivalently, $Y^{\ast}_{t,k} \sim \mathcal{N} (Z^{\ast}_{t,k}, \sigma^2_{\text{me},k} )$.

At each MCMC iteration, posterior samples of the latent “true" components were transformed back to the original pollutant scale as we aim to yield interpretable estimates from the outcome model.
We elaborate on this further in the next section.
\subsection{Back Transformation and Outcome Poisson Model}\label{Approach 1}

\begin{figure}[!t]
  \centering
  \includegraphics[width=0.9\linewidth]
  {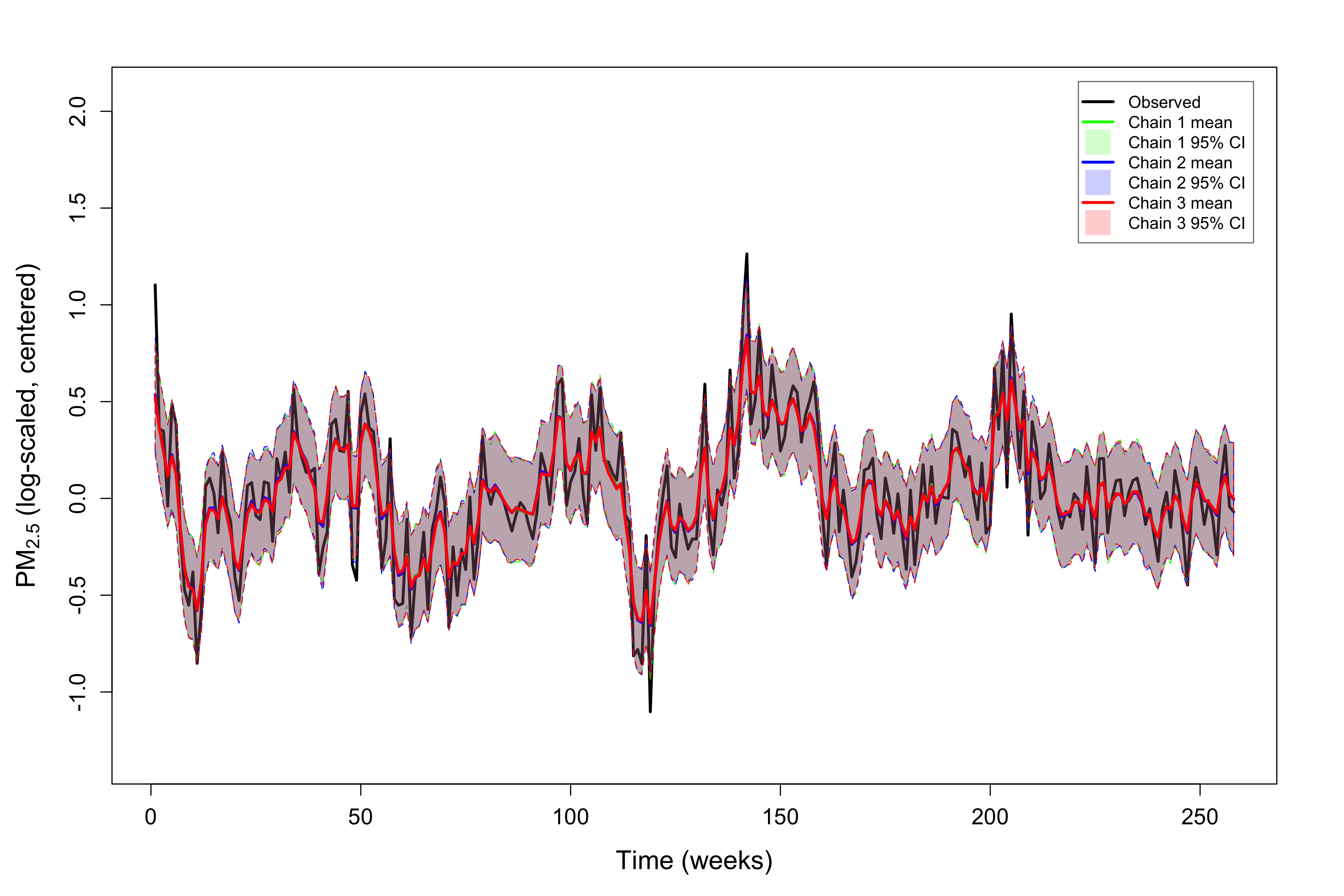}
  \caption{Comparison of   PM$_{\text{2.5}}$ pollution levels and estimated true exposure values on the log-transformed and centred scale. Observed measurements are shown in black. Estimated true exposures are presented separately for three MCMC chains: Chain 1 (green), Chain 2 (blue), and Chain 3 (red). For each chain, the solid line represents the posterior mean, and the corresponding dashed lines delimit the shaded credible intervals, with interval fill matching the chain's color.}
  \label{fig:estimated pollution level}
\end{figure}
We back-transformed $Z^{\ast}_{t,k}$ using the orthogonal transformation matrix $Q \in \mathbb{R}^{K \times K}$.
Denoting $Z_t^* \in \mathbb{R}^{1 \times K}$ as the row vector of latent levels in the transformed space, the latent levels in the original space are $Z_t = Z_t^* Q^\top \in \mathbb{R}^{1 \times K}$.
Equivalently, $Z_{t,j} = \sum_{k=1}^K Q_{j,k} \, Z^{\ast}_{t,k}$,
where $Z_{t,j}$ is the latent true level on the original scale of pollutant  $j$ at week $t$. These back-transformed values were used as predictors in the Poisson outcome model. 
When applied to our data, the three MCMC chains are leading to very similar exposure estimates; the estimated exposure levels differ little from the observed values. Figure \ref{fig:estimated pollution level} is an example for the particulate matter pollutant PM\textsubscript{2.5}. The plots for the other pollutants are shown in the \textbf{Supplementary Materials} Figure S1.

In the outcome model, $M_t$ denoted the observed mortality count at week $t\in\{2,...,N\}$ and follows the $\text{Poisson}(\lambda_t)$ distribution, where $\lambda_t$ is the expected mortality count at time t and 
\[ \log (\lambda_t) = \alpha + \sum_{k \in K'} \beta_k Z_{t-1,k} + \gamma T_{t-1} + \log(O_t),
\]
where $Z_{t-1,k}$ refers to the true pollution level of pollutant k at time $t-1\in\{1,..., N-1\}$, $\alpha$ is the intercept, $\beta_k$ is the effect of lag-1 pollutant $k$, $\gamma$ is the lag-1 temperature effect, $T_t$ is the temperature at week $t$ ($T_t$ may be expressed either on the original scale, or as a threshold-adjusted value, or be completely removed depending on the model specification), and $O_t$ is the population within the county at risk at time $t$. Here, \( K' \subseteq \{1,2,3,4,5,6\} \) denotes the subset of indices corresponding to the selected pollutants, as mentioned in Section \ref{SVD Transformation}. As will discuss in Section~\ref{Model Selection}, not all six pollutants are necessarily included in the final model.

Simulation studies of the proposed model framework were all conducted in \texttt{R} to evaluate the model's performance and robustness. We demonstrate this in more detail in the next section.

\section{Simulation Studies}\label{simulation}
\subsection{Simulation}\label{Simulation Structure}

To evaluate the proposed modeling framework, six independent air pollution time series, denoted as $Z_k, k \in \{1,\dots,6\}$, each comprising $N = 250$ time points, were first simulated to represent the true pollution levels in the SVD-transformed space. Each series $Z_k$ was generated as an AR(1) process:$Z_{k,t} = \phi_k Z_{k,t-1} + \eta_{k,t}$ where $\eta_{k,t} \sim \mathcal{N}(0, \tau^2_k)$
where $|\phi_k| < 1$ ensures stationarity. $\eta_{k,t}$ captured the random fluctuations in the true pollution process, with variance $\tau^2_k$ controlling the magnitude of these fluctuations. To initialize the series under stationarity, the first value $Z_{k,1}$ was drawn from the stationary distribution: $Z_{k,1} \sim \mathcal{N}\left(0, \tau^2_k/(1 - \phi_k^2)\right)$, which ensures that the variance of $Z_{k,t}$ remains constant over time. Measurement error was then introduced to generate the transformed observed pollution values, $Y_{k,t}$, as follows $Y_{k,t} = Z_{k,t} + \epsilon_{k,t}$, where $\epsilon_{k,t} \sim \mathcal{N}(0, \sigma^2_k)$,
with $\epsilon_{k,t}$ independent of $\eta_{k,s}$ for all $t, s$. Here, $\epsilon_{k,t}$ denotes the additive measurement error, with variance $\sigma^2_k$ governing the noise level for the $k$-th series. The temperature variable $T_t$ was modeled as uniformly distributed between 5 and 30 to approximate the real data.

Using the simulated pollution series, health outcome counts were generated from a Poisson log-linear model in which the expected number of cases depended on the pollution levels and temperature. Fixed true coefficients were specified for the pollution effects, the temperature effect, and the intercept 
($\beta_1=0.1, 
  \beta_2=-0.2,  \beta_3=0.3
  ,  \beta_4=-0.15,  \beta_5=0.25,  \beta_6=-0.1, \gamma=-0.01\text{ and } , \alpha =-12  $). These expected counts were then used to simulate Poisson-distributed health outcomes: $M_t \sim \text{Poisson}(\lambda_t)$, where for $K\equiv \{1,2,3,4,5,6\}$
\[ \log (\lambda_t) = \alpha + \sum_{k\in K} \beta_k Z_{t-1,k} + \gamma T_{t-1}.
\]

For inference, we implemented the full Bayesian hierarchical model in Just Another Gibbs Sampler (JAGS) via the “rjags" package in R. The model jointly estimated the latent pollution states, autoregressive parameters, and regression coefficients linking pollution and temperature to the health outcome.  Specifically, let $Z_{t,k}$ be the latent true pollution level, and $Y_{t,k}$ be the observed pollution level. The latent pollution process in the rotated space follows an AR(1) structure:
$Z_{t,k} =
\phi_k Z_{t-1,k}
+
\epsilon_{t,k}$
where
$\epsilon_{t,k} \sim N(0,\tau_{proc,k}^{-1})$.
The noisy measurements of the true latent process in the rotated space was $Y^*_{t,k} \sim N(Z_{t,k}, \tau_{me,k}^{-1})$.
For $t = 1,\dots,N$, the expected health outcome count was modeled as $M_t \sim \text{Poisson}(\lambda_t)$ with
\[
\log(\lambda_t)
=
\alpha +
\sum_{k\in K} \beta_k Z_{t-1,k}
+
\gamma T_{t-1}
,
\]
where $M_t$ represents the mortality counts and $T_t$ is the observed temperature. Noninformative priors were assigned to all parameters, including regression coefficients and variance components. 

For the MCMC implementation, three MCMC chains were run in parallel, with a burn-in period of 10,000 iterations discarded to eliminate the influence of initial values, followed by 3,000 post-burn-in iterations for posterior sampling. To assess estimation stability, the entire simulation and fitting procedure was replicated 1,000 times, and the resulting parameter estimates and credible intervals were recorded across all repetitions.

We give only a brief description of the simulation results here; extensive details about the simulation results are shown in \textbf{Supplementary Material} Section 3 and Figure S2. Overall, the results demonstrate that model performance and parameter recovery were optimal when the inferential model matched the true data-generating process, whether that involves measurement error, the full set of pollutants, or the correct pollution variable (latent vs.~observed). The true data-generating model produced the lowest WAIC.   The results confirm that the latent time series and measurement error modelling strategy is readily implementable at scale.

\section{Application to the Los Angeles County Data} \label{Real Data Analysis}

\subsection{Descriptive Data Analysis} \label{sec:DescData}
\begin{figure}[!t]
  \centering
  \includegraphics[width=0.9\linewidth]
  {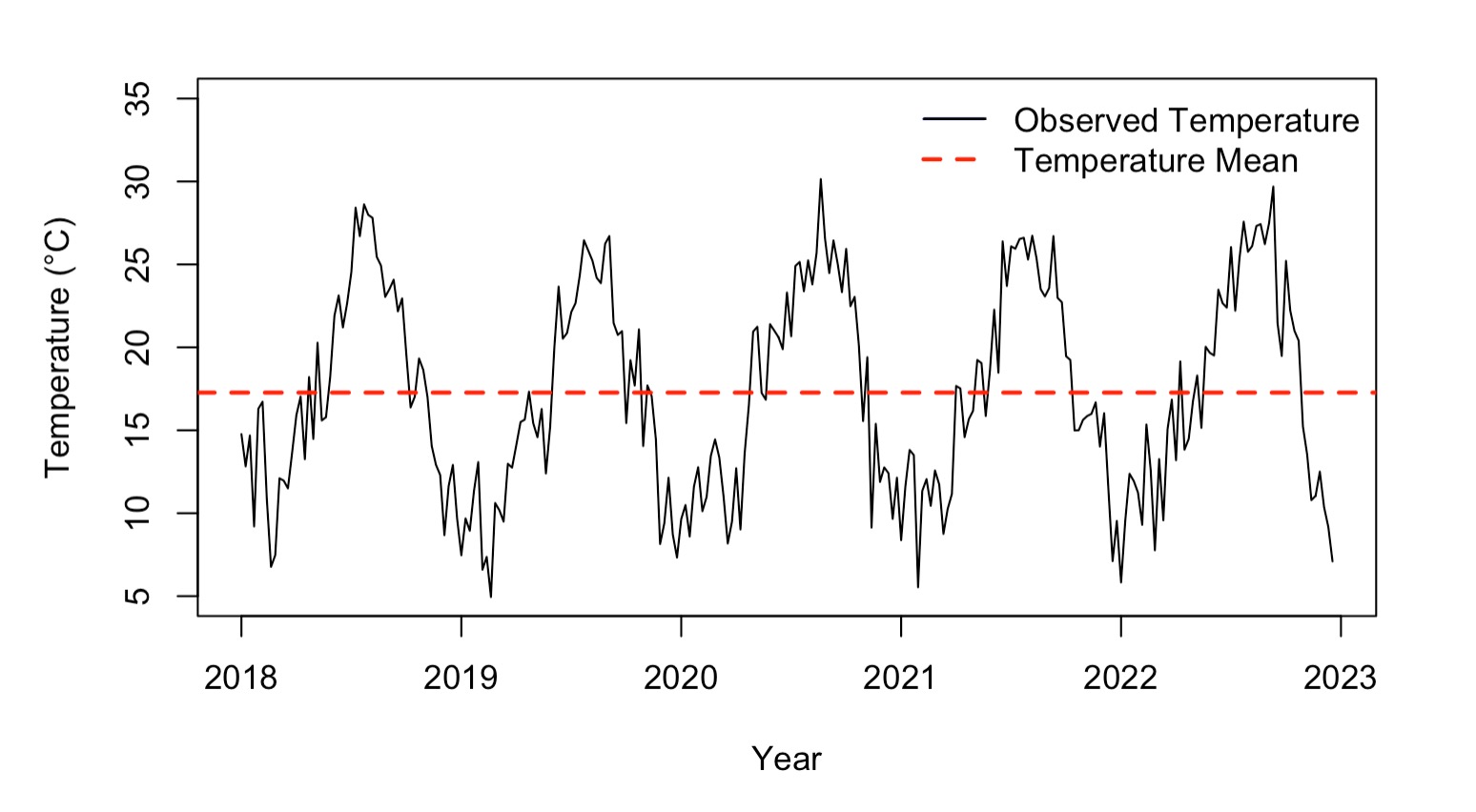}
  \caption{Temperature time series in LA County (January 1, 2018 - December 17, 2022)}
  \label{fig:temperature}
\end{figure}

Table~\ref{tab:mean} summarizes the distribution of death counts, temperature, and air pollutant levels in LA County over the study period from January 1, 2018, to December 17, 2022, covering 259 MMWR weeks. Mean weekly concentrations of major pollutants were as follows: PM$_{2.5}$ at 11.57~$\mu$g/m$^3$, PM$_{10}$ at 26.45~$\mu$g/m$^3$, SO$_2$ at 0.36~ppb, NO$_2$ at 13.89~ppb, CO at 0.36~ppm, and O$_3$ at 0.03~ppm. Air pollution time series plots are shown in Figures~\ref{fig:PM2.5}-\ref{fig:O3}. 
According to the \href{https://ww2.arb.ca.gov/resources/california-ambient-air-quality-standards}{California Ambient Air Quality Standards}, the standard levels for the six pollutants in LA County are 35~$\mu$g/m$^3$, 50~$\mu$g/m$^3$, 0.04~ppm (40~ppb), 180~ppm (180,000~ppb), 9~ppm, and 0.07~ppm, respectively. Comparing these standards with Table~\ref{tab:mean}, pollutant levels were generally below the standards, except for PM$_{10}$, which exhibited a few peaks exceeding the standard level. 
Temperature (Figure~\ref{fig:temperature}) averaged 17.28~$^\circ$C, ranging from 4.95~$^\circ$C to 30.14~$^\circ$C. As seen in Figures~\ref{fig:PM2.5}-\ref{fig:O3},~\ref{fig:temperature}, and Table~\ref{tab: Correlation Matrix}, temperature was negatively associated with CO and NO$_2$ concentrations, which declined during warmer months. In contrast, O$_3$ (Figure~\ref{fig:O3}) closely tracked temperature, with elevated concentrations during hot months. SO$_2$ levels (Figure~\ref{fig:SO2}) decreased from 2018 to 2019, plateaued, and then rose again after 2019. PM$_{2.5}$ and PM$_{10}$ remained relatively stable overall but exhibited higher winter peaks during 2020--2021. Table~\ref{tab: Correlation Matrix} presents the correlation matrix, showing relationships among pollutants and temperature across this five-year span.

\begin{figure}[ht]
    \setkeys{Gin}{width=1\linewidth,height=0.15\textheight}
    \centering
    \begin{minipage}{0.45\linewidth}
        \centering
        \subfloat[COPD]{\includegraphics{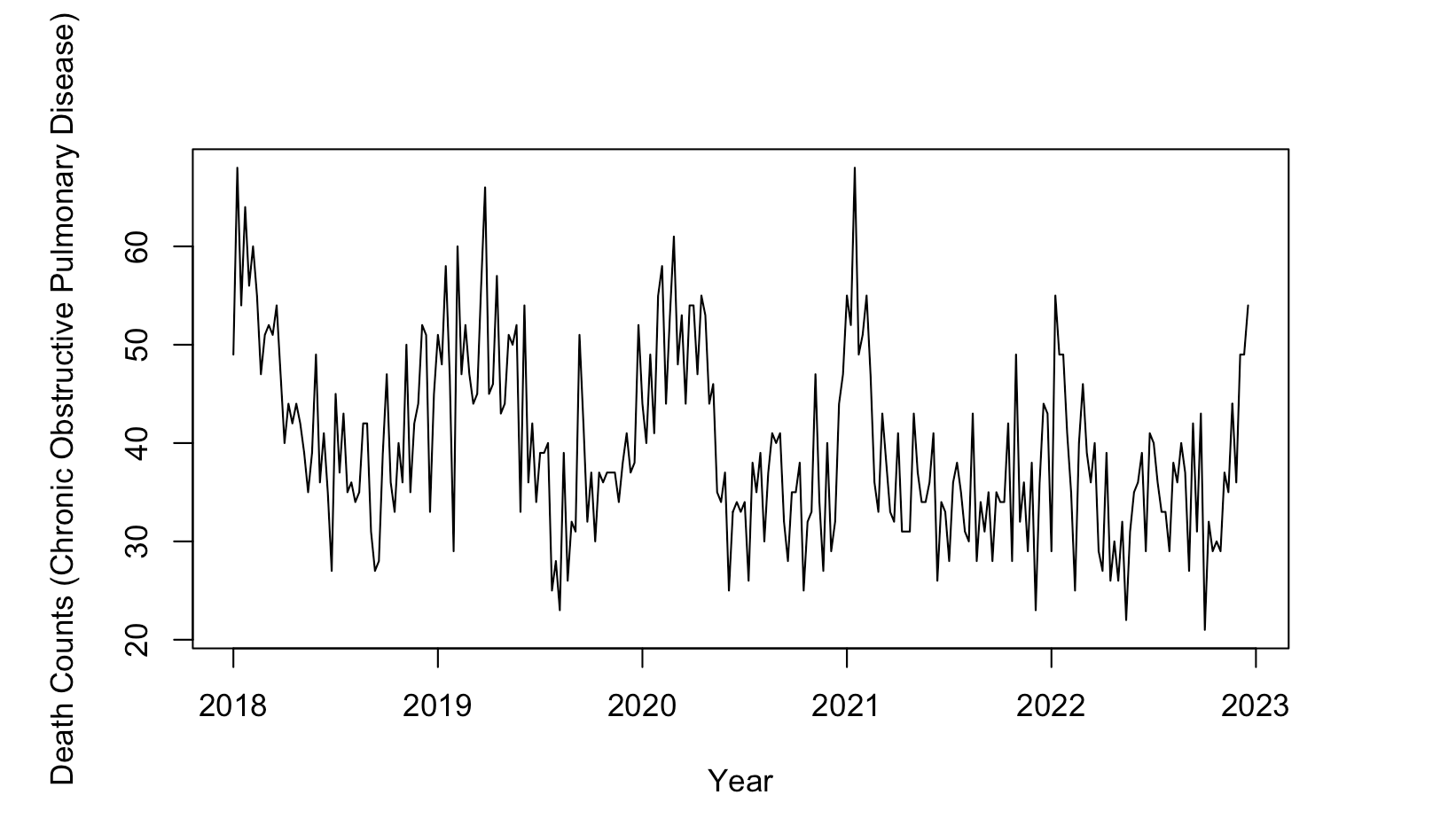}\label{fig:COPD}}
    \end{minipage}
    \quad
    \begin{minipage}{0.45\linewidth}
        \centering
        \subfloat[Pneumonia]{\includegraphics{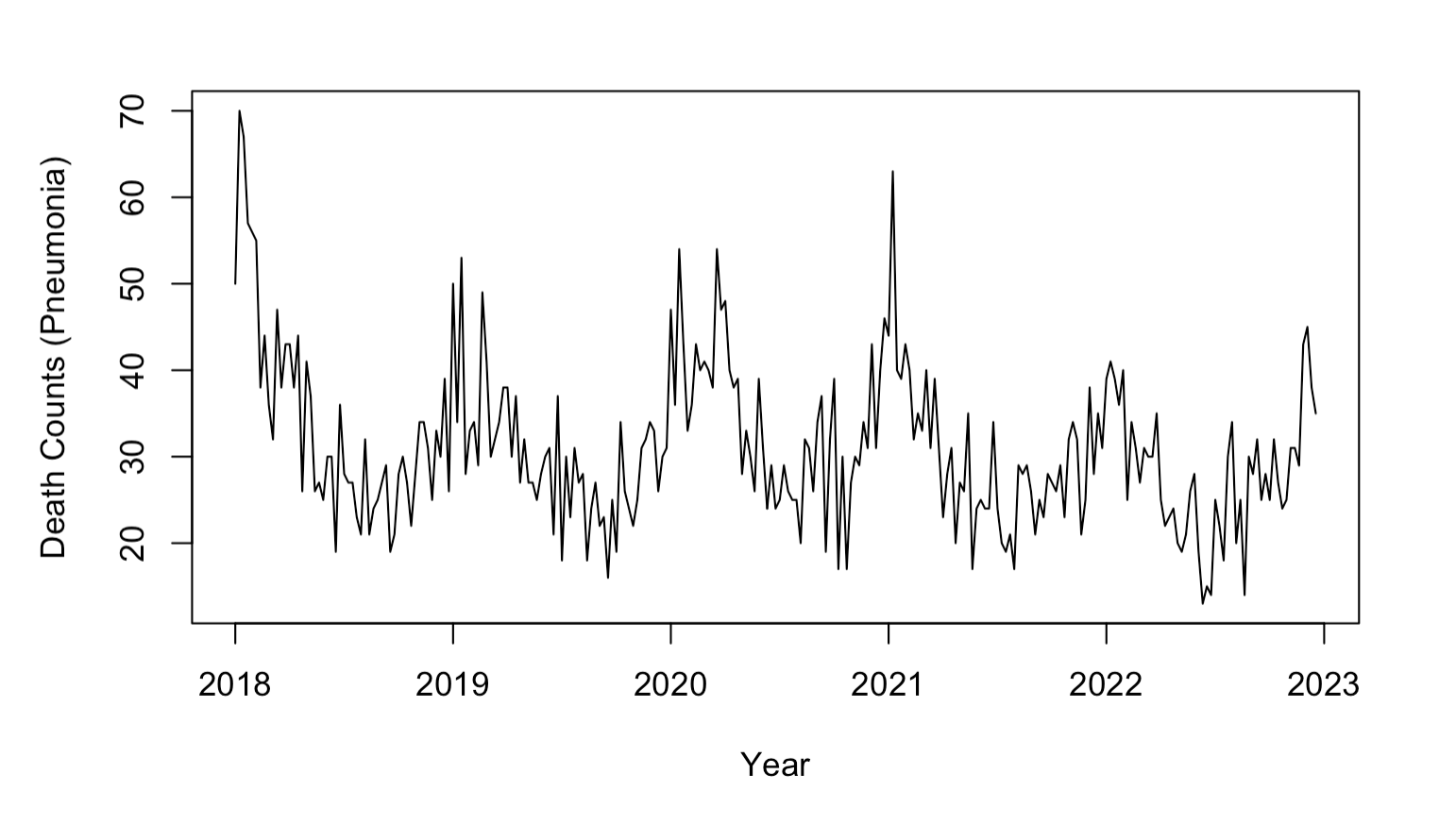}\label{fig:Pneumonia}}
    \end{minipage}
    
    \vspace{1ex}
    
    \begin{minipage}{0.45\linewidth}
        \centering
        \subfloat[Heart Failure]{\includegraphics{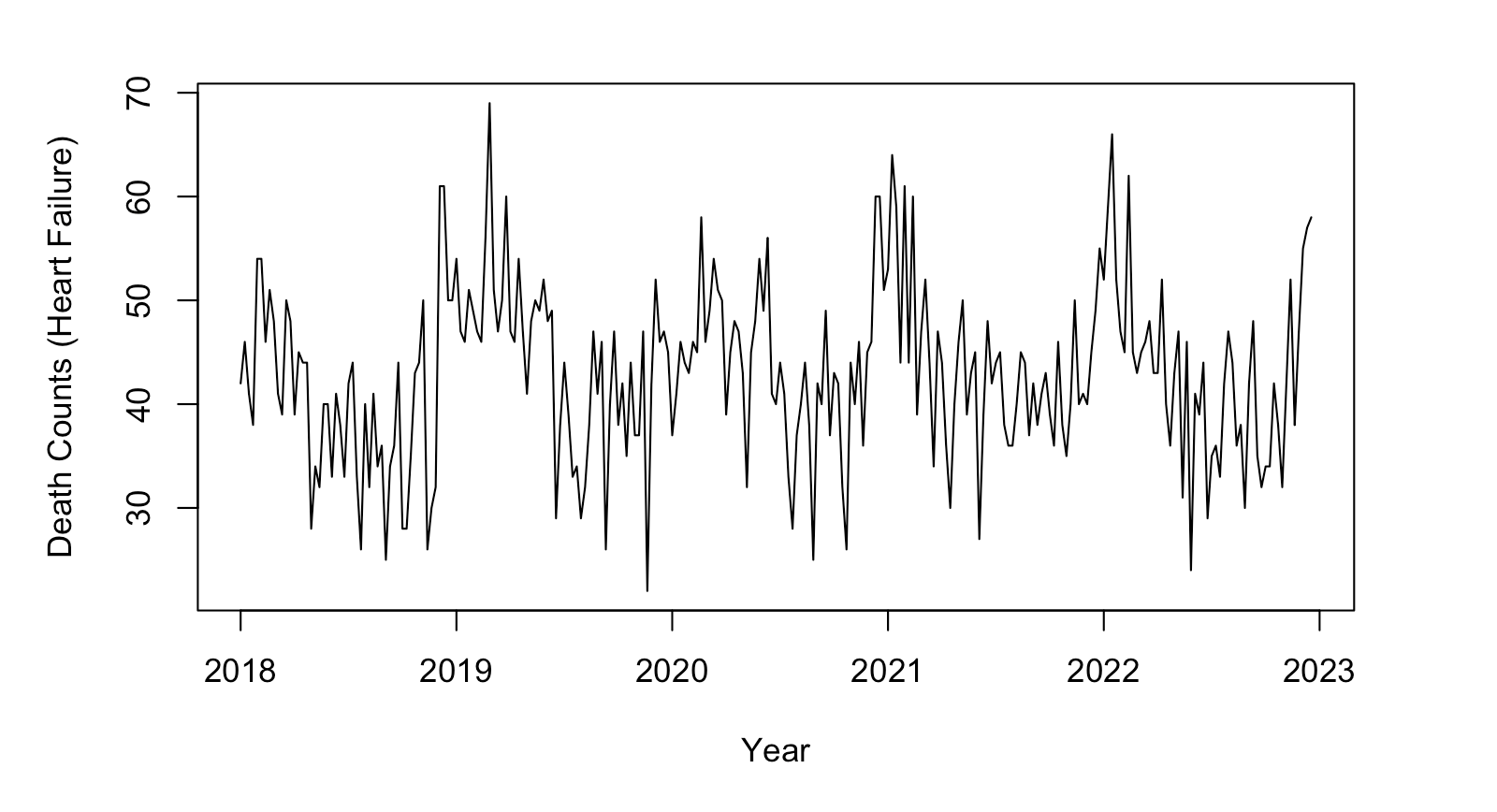}\label{fig:Heart Failure}}
    \end{minipage}
    \quad
    \begin{minipage}{0.45\linewidth}
        \centering
        \subfloat[Neoplasm]{\includegraphics{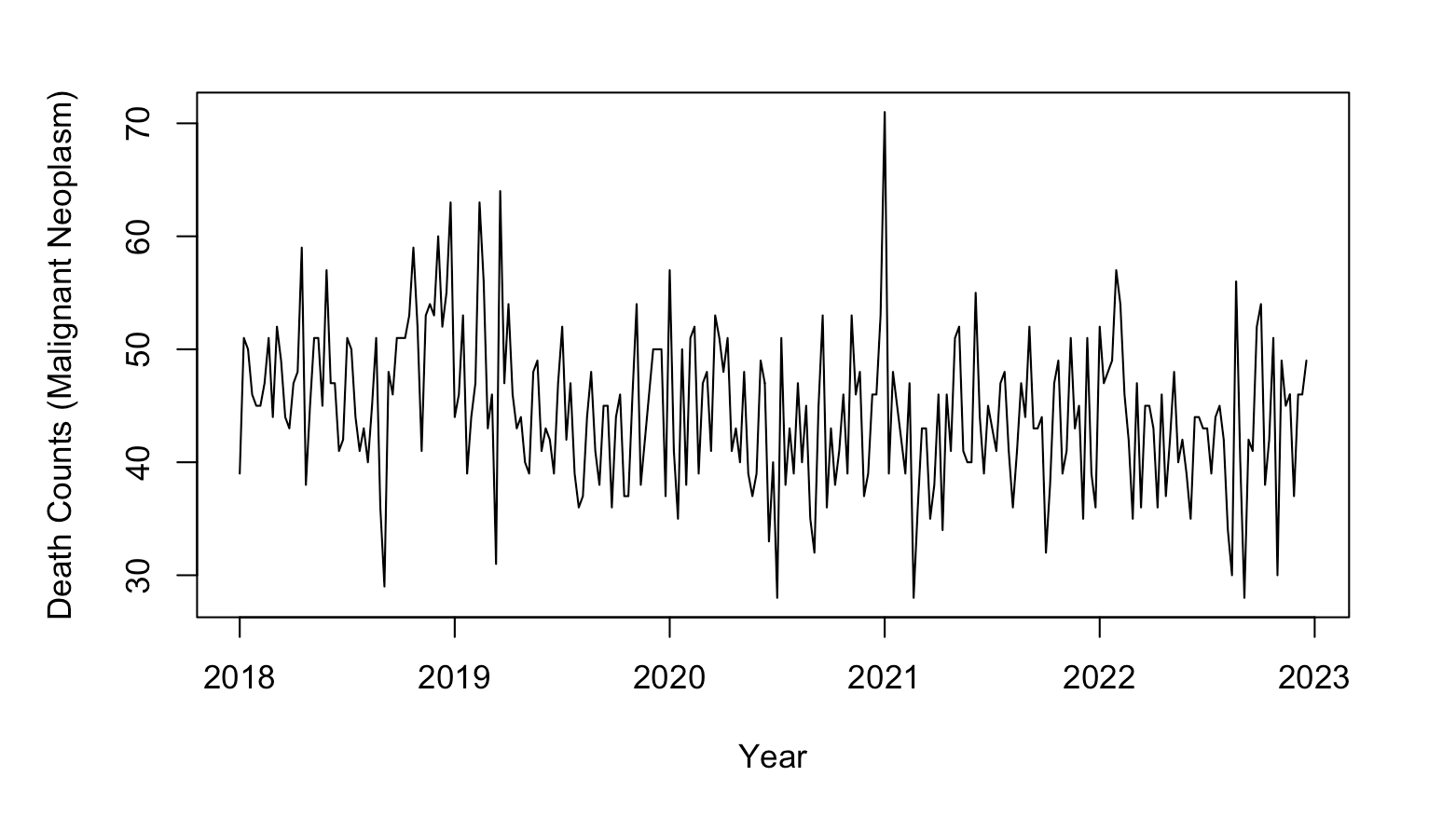}\label{fig:Neoplasm}}
    \end{minipage}
    
    \vspace{1ex}
  
    \caption{Time series of four health outcomes in LA County from January 1, 2018, to December 17, 2022.}
    \label{fig:main}
\end{figure}

Total mortality from the four causes over the five years reached 40,982 cases: 10,287 from COPD, 8,038 from pneumonia, 11,541 from neoplasms, and 11,116 from heart failure, with time series plots shown in Figures ~\ref{fig:COPD}-\ref{fig:Neoplasm}. For all four causes, there were higher peaks observed during colder months. The time series for neoplasms (Figure~\ref{fig:Neoplasm}) shows minimal structure, while all other causes Figures~\ref{fig:COPD},~\ref{fig:Pneumonia}, and~\ref{fig:Heart Failure}) exhibit seasonality, with peaks at the end of the year.

\captionsetup[table]{skip=8pt}
\begin{table*}[!t]
\centering
\footnotesize
\caption{Summary statistics of weekly variables (January 1, 2018 - December 17, 2022).}
\begin{tabularx}{\linewidth}{@{}ll*{3}{>{\centering\arraybackslash}X}*{5}{>{\centering\arraybackslash}X}@{}}
\toprule
\multirow{2}{*}{} & \multirow{2}{*}{} & \multirow{2}{*}{\textbf{Mean}} & 
\multirow{2}{*}{\textbf{SD} ($\sigma$)} 
& \multicolumn{5}{c}{\textbf{Percentage}} \\
\cmidrule(lr){5-9}  &  &  &  & \textbf{Min} & \textbf{25} & \textbf{Median} & \textbf{75} & \textbf{Max} \\
\midrule
\multirow{6}{*}{\begin{tabular}{@{}l@{}} \textbf{Air pollutants} \\ \textbf{concentration} \end{tabular}} 
 & PM$_{2.5}$ ($\mu$g/m$^3$) & 11.57 & 4.15 & 3.54 & 8.99 & 10.83 & 13.20 & 37.69 \\
 & PM$_{10}$ ($\mu$g/m$^3$) & 26.45 & 10.42 & 3.79 & 20.58 & 26.69 & 31.66 & 87.34 \\
 & SO$_{2}$ ($\text{ppb}$) & 0.36 & 0.17 & 0.01 & 0.24 & 0.34 & 0.46 & 1.03 \\
 & NO$_{2}$ ($\text{ppb}$) & 13.89 & 5.08 & 5.68 & 9.77 & 12.82 & 16.22 & 29.56 \\
 & CO ($\text{ppm}$) & 0.36 & 0.13 & 0.17 & 0.25 & 0.32 & 0.43 & 0.78 \\
 & O$_{3}$ ($\text{ppm}$) & 0.03 & 0.01 & 0.01 & 0.02 & 0.03 & 0.04 & 0.05 \\
\midrule
\multirow{1}{*}{\begin{tabular}{@{}l@{}} \textbf{Temperature} \end{tabular}} 
 & Temperature ($^\circ$C) & 17.28 & 6.13 & 4.95 & 12.32 & 16.61 & 22.66 & 30.14 \\
\midrule
\multirow{4}{*}{\begin{tabular}{@{}l@{}} \textbf{Weekly deaths,} \\ \textbf{count} \end{tabular}} 
 & COPD & 40 & 9.23 & 21.00 & 33.00 & 38.00 & 45.75 & 68.00 \\
 & Pneumonia & 31 & 9.11 & 13.00 & 25.00 & 30.00 & 35.75 & 70.00 \\
 & Neoplasm & 45 & 6.72 & 28.00 & 40.00 & 45.00 & 49.00 & 71.00 \\
 & Heart Failure & 43 & 8.32 & 22.00 & 38.00 & 43.00 & 47.00 & 69.00 \\
\bottomrule
\end{tabularx}

\begin{minipage}{\linewidth}
\footnotesize
1~$\mu$g/m$^3$, 1 microgram (10\textsuperscript{-6} grams) of pollutant per cubic meter of air; 1 ppm: 1 milligrams per liter (mg/L); 1 ppb, 1 microgram per liter.
\end{minipage}
\label{tab:mean}
\end{table*}

\begin{table}[h]
\centering
\caption{Correlation matrix of the predictors (January 1, 2018 - December 17, 2022).}
\begin{tabular}{lrrrrrrr}
\toprule
                   & PM\textsubscript{2.5} & PM\textsubscript{10} & SO\textsubscript{2} & NO\textsubscript{2} & CO    & O\textsubscript{3} & \multicolumn{1}{c}{Temperature} \\
\midrule
PM\textsubscript{2.5} & 1.000 & 0.583 & 0.181 & 0.561 & 0.677 & -0.187 & 0.130 \\
PM\textsubscript{10}  & 0.583 & 1.000 & 0.195 & 0.037 & 0.066 & 0.412  & 0.639 \\
SO\textsubscript{2}   & 0.181 & 0.195 & 1.000 & 0.367 & 0.257 & -0.075 & 0.231 \\
NO\textsubscript{2}   & 0.561 & 0.037 & 0.367 & 1.000 & 0.952 & -0.728 & -0.400 \\
CO                    & 0.677 & 0.066 & 0.257 & 0.952 & 1.000 & -0.716 & -0.405 \\
O\textsubscript{3}    & -0.187 & 0.412 & -0.075 & -0.728 & -0.716 & 1.000 & 0.670 \\
Temp.           & 0.130 & 0.639 & 0.231 & -0.400 & -0.405 & 0.670  & 1.000 \\
\bottomrule
\end{tabular}

\label{tab: Correlation Matrix}
\end{table}

\subsection{Prior Distributions in the Bayesian Model}
Based on the model framework outlined in Section~\ref{Method}, we assigned prior distributions to all unknown parameters within the Bayesian hierarchy. While conventional noninformative priors with large variances are commonly used, they can often lead to numerical instability. To mitigate this, we adopted weakly informative priors that constrain parameters to practically plausible ranges.

As a preliminary step, we fitted a conventional Poisson model using the standard \texttt{glm} function in \textsf{R} with \texttt{family = Poisson}, including all pollution variables as predictors, population as the offset, and the health outcome as the response, while excluding the state-space component. The resulting coefficient estimates for all pollution-related predictors were very close to zero. Consequently, setting the prior mean of $\beta_k$ to the GLM estimate $\hat{\beta}_k$ was effectively equivalent to centering it at zero, making little practical difference.

For the precision parameters governing process variability and measurement error in each latent component, we specified independent Gamma distributions: $\tau_{\text{proc},k} \sim \text{Gamma}(0.5,0.5)$, $\tau_{\text{me},k} \sim \text{Gamma}(0.5,0.5)$.
Regression coefficients linking the latent components to the intermediate and outcome models—namely $a_k$, $b_k$, $\beta_{1k}$, $\beta_{2k}$, $\beta_k$, and $\alpha_k$—were each assigned independent zero-mean Normal priors with small precision: $\mathcal{N}(0,0.01)$.
For temperature-related variability, we placed a noninformative Gamma prior on the precision parameter: $\tau_{\text{temp}} \sim \text{Gamma}(0.01,0.01)$.
Finally, the elements of the latent interaction matrix $\phi_k$ were each assigned independent Normal priors.

\subsection{Results: Model Selection and Elaboration}\label{Model Selection}
Due to the large number of potential covariate combinations arising from the six predictor pollutants and the observed temperature, 
an exhaustive search of all plausible models would have been computationally infeasible. For this reason, we applied methods of model selection similar to the ones used in the frequentist paradigm, that is, forward selection, backward elimination, and two-way stepwise elimination. These heuristics provide a structured and efficient way of exploring the model space by adding and/or removing covariates from it, thereby selecting only a few possible models. The Akaike information criterion (AIC) was replaced with the Watanabe--Akaike information criterion (WAIC), which is more appropriate in the context of the Bayesian approach since it provides an estimate of out-of-sample prediction accuracy while taking into account the complexity of the model. Each method resulted in a candidate model, and the selected model for inference was the one with the smallest WAIC.

Although stepwise procedures are inherently path-dependent and may lead to different selected models, the use of WAIC could provide a principled and consistent basis for comparison across candidate models. By combining results from the three path searches, the approach enhanced robustness by ensuring that conclusions are not driven by a single search path. While it did not necessarily return identification of the globally optimal model, it provided a practical and theoretically grounded approach for identifying a model with strong predictive performance, making it suitable for applied Bayesian analysis in high-dimensional settings.


To determine the most appropriate model configuration for capturing the relationships of interest, we also considered an alternative specification that excludes PM\textsubscript{10} from the selected model, but only in cases where both PM\textsubscript{2.5} and PM\textsubscript{10} were retained in the model selected through the procedure described in Section \ref{Model Selection}. For each such outcome, we removed PM\textsubscript{10} and compared the WAIC of this reduced model against that of the originally selected model. This approach was motivated by the reasoning that PM\textsubscript{10} may not contribute substantial independent information beyond PM\textsubscript{2.5}. As noted by Cheng \citep{cheng2015pm2} and discussed in the \href{https://ww2.arb.ca.gov/resources/inhalable-particulate-matter-and-health}{California Air Resources Board report on inhalable particulate matter and health}, PM\textsubscript{2.5} accounts for approximately 70\% of PM\textsubscript{10}. Consequently,
including both pollutants may offer little additional independent information beyond what is already captured by PM\textsubscript{2.5} alone.

For each selected outcome in which temperature was not eliminated during the selection process, raw temperature was retained as one of the main predictors. To extend our analysis, we then compared this model against two alternatives using the WAIC: (ii) threshold-adjusted temperature, and (iii) exclusion of temperature. For approach (ii), we selected 18\,$^\circ$C as the cut point in our analysis; previous research \citep{gomez2013cold} has established that mortality, particularly from cardiovascular and respiratory causes, is associated with cold temperatures, though such effects are typically observed under more extreme conditions, such as temperatures below 0\,$^\circ$C. Given the relatively mild temperature range in Los Angeles County (approximately 4.95\,$^\circ$C to 30.14\,$^\circ$C), extreme cold is uncommon; however, we were interested in examining how relatively cold temperatures within this range might affect mortality risk. This selected cutoff closely corresponded to the mean temperature: 17.28$^\circ$C observed in our dataset for Los Angeles County.

For the threshold-adjusted temperature specification, we transformed the temperature variable as follows: for values above 18\,$^\circ$C, the temperature was set to 0; for values below 18\,$^\circ$C, we subtracted 18 from the original value. This coding scheme allowed the intercept to be interpreted as the effect at the threshold temperature, while the coefficient for the adjusted variable captured the additional effect of temperatures above the threshold. This approach enabled us to isolate and examine the contribution of the lower portion of the temperature distribution to mortality risk.

\subsection{Results for the Four Outcome Measures} \label{Results}
\begin{figure}[ht]
    \setkeys{Gin}{width=1\linewidth,height=0.15\textheight}
    \centering
    \begin{minipage}{0.45\linewidth}
        \centering
        \subfloat[PM\textsubscript{2.5}]{\includegraphics{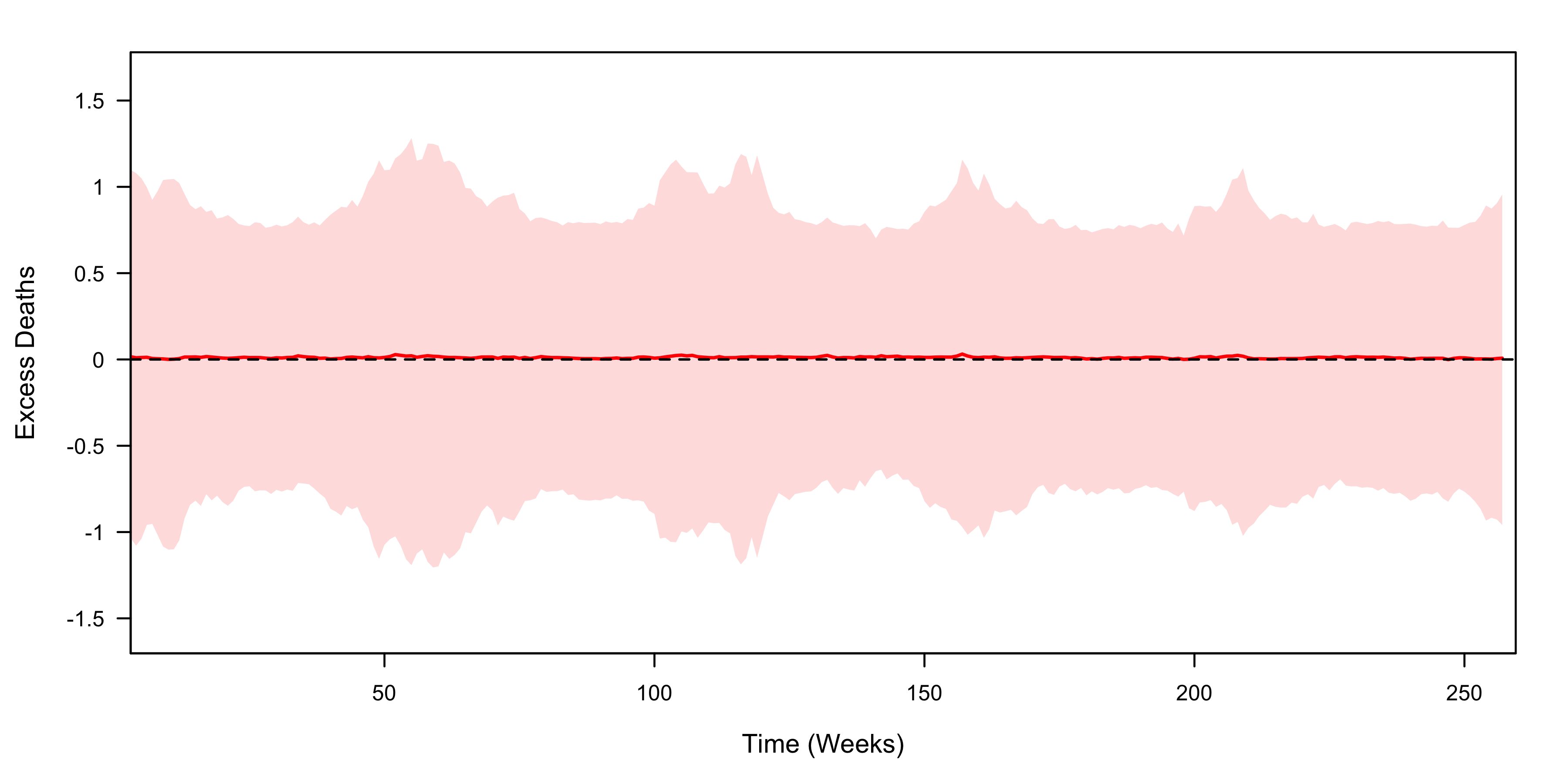}\label{fig:COPD-PM10}}
    \end{minipage}
    \quad
    \begin{minipage}{0.45\linewidth}
        \centering
        \subfloat[PM\textsubscript{10}]{\includegraphics{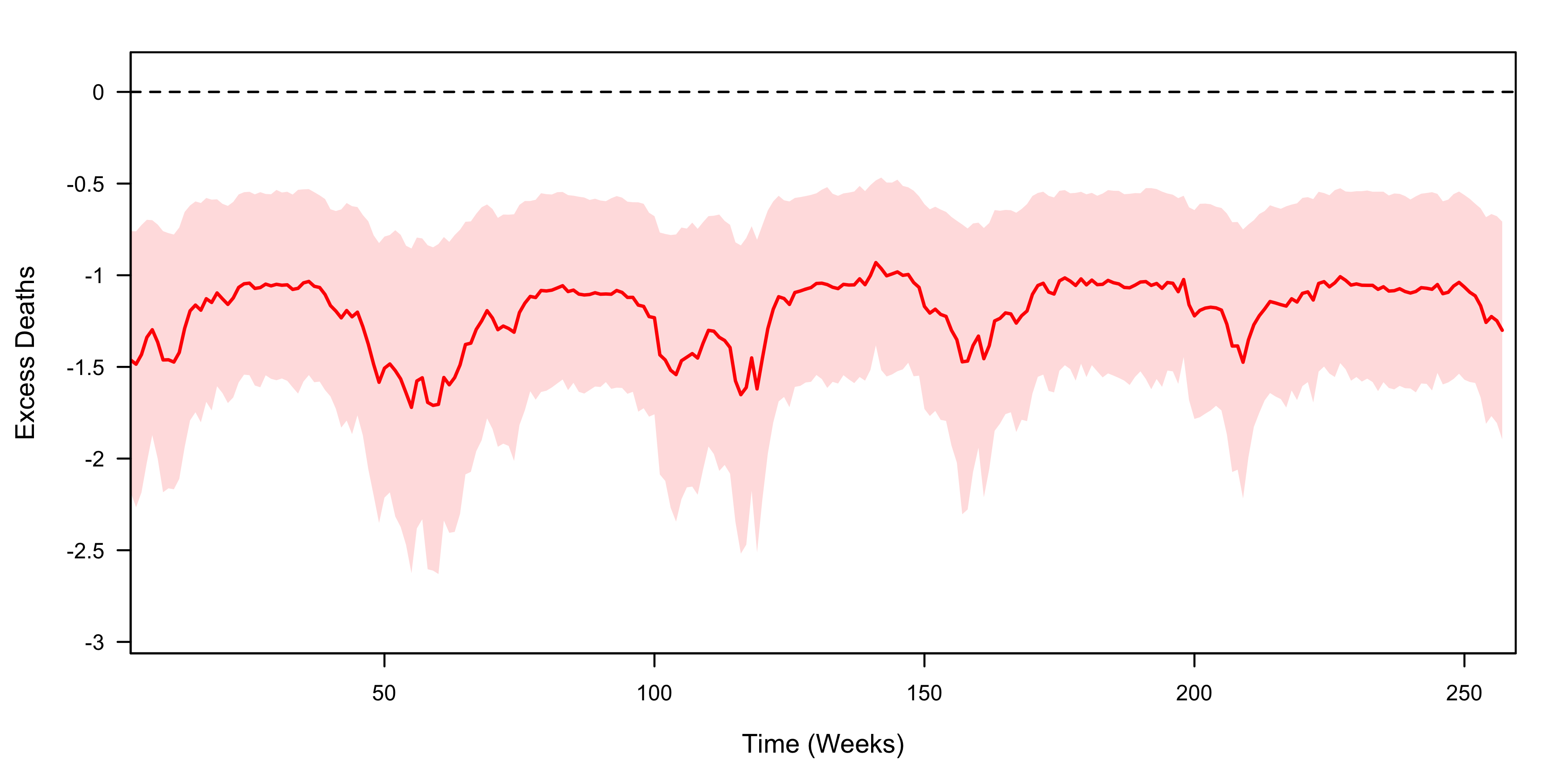}\label{fig:COPD-CO}}
    \end{minipage}
    
    \vspace{1ex}
    
    \begin{minipage}{0.45\linewidth}
        \centering
        \subfloat[CO]{\includegraphics{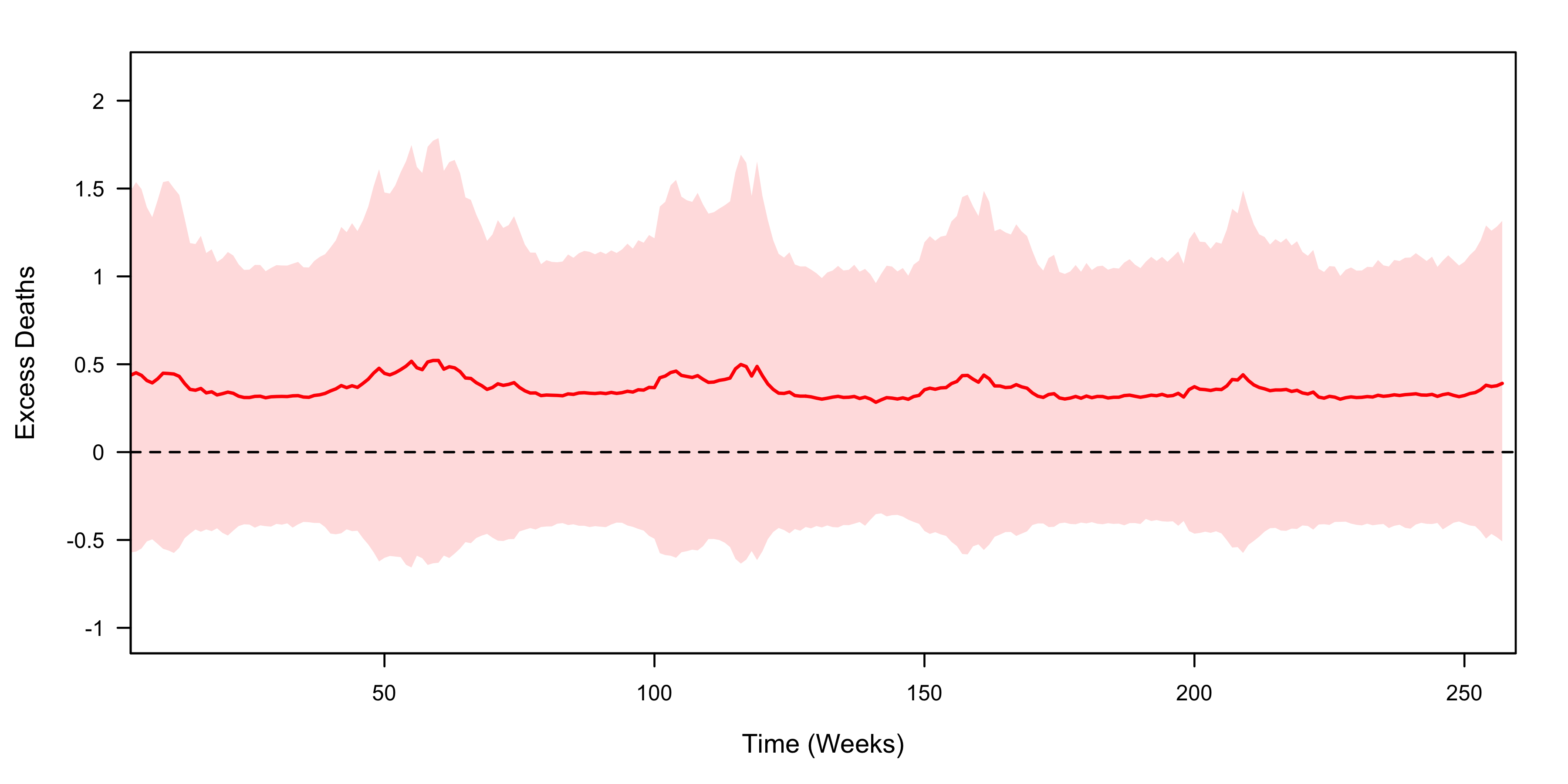}\label{fig:COPD-O3}}
    \end{minipage}
    \quad
    \begin{minipage}{0.45\linewidth}
        \centering
        \subfloat[O\textsubscript{3}]{\includegraphics{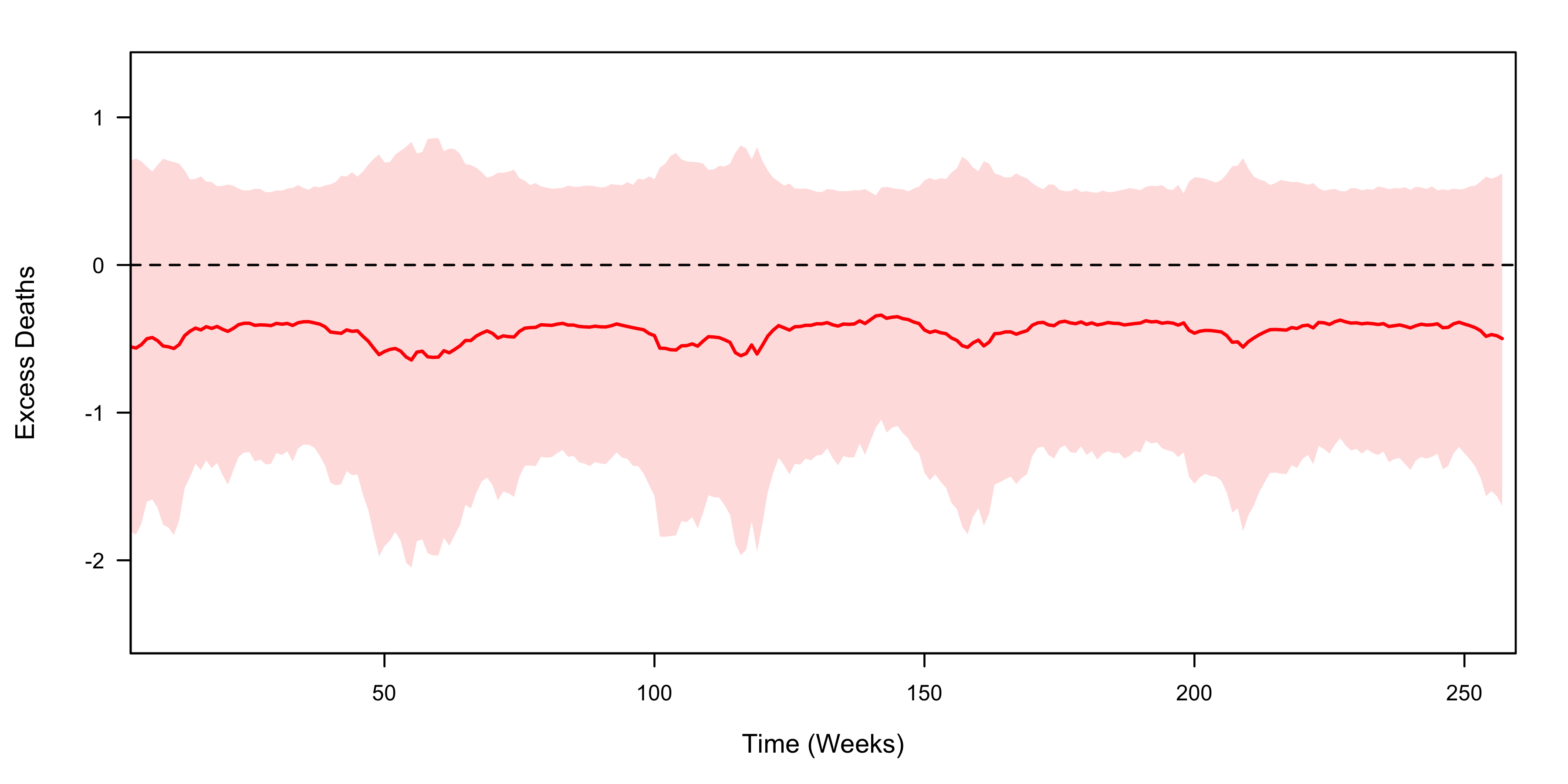}\label{fig:COPD-O3}}
    \end{minipage}
    
    \begin{minipage}{0.45\linewidth}
        \centering
        \subfloat[Temperature]{\includegraphics{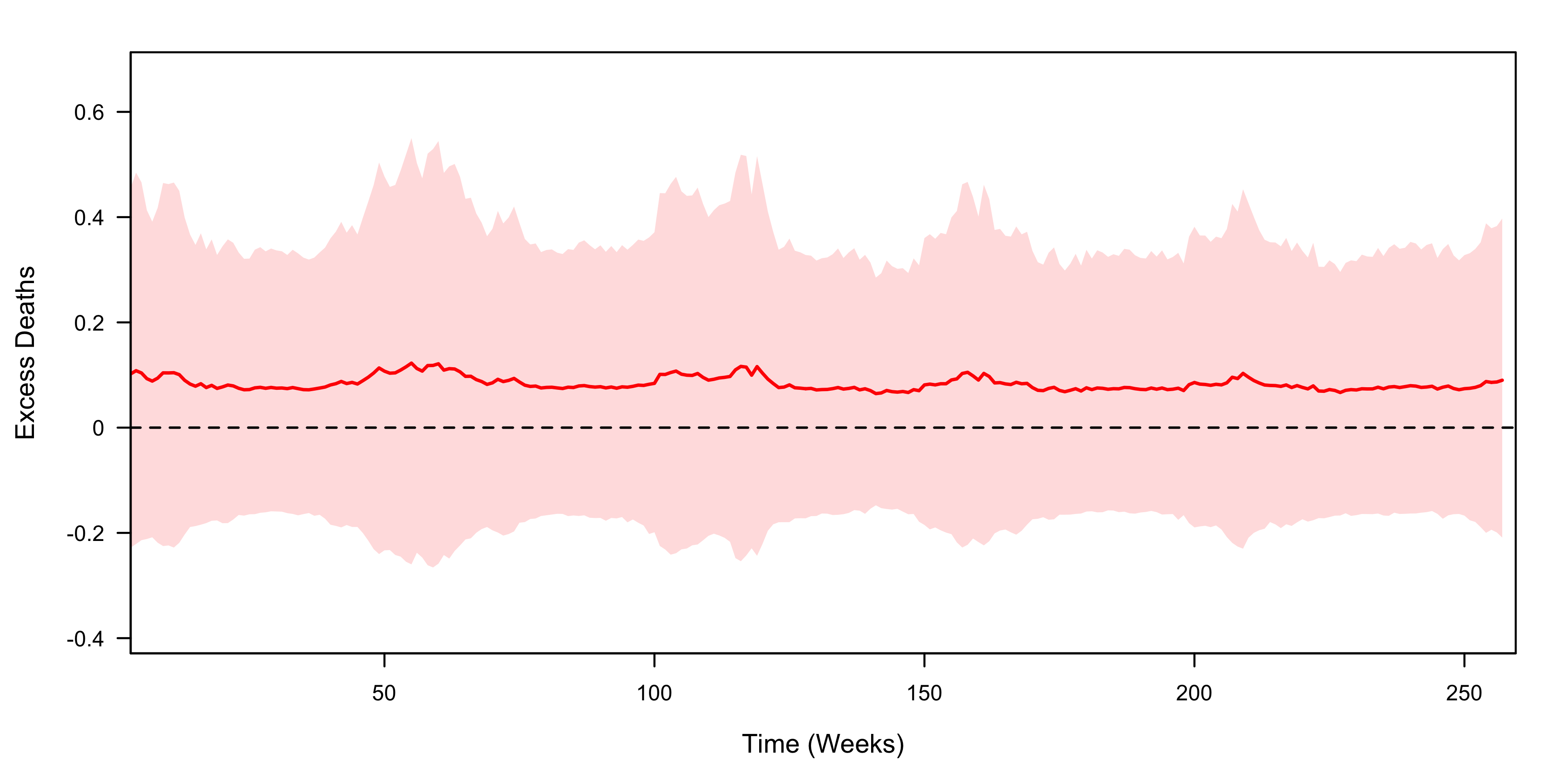}\label{fig:COPD-Temperature}}
    \end{minipage}
    
    \caption{Averaged estimated changes in weekly COPD mortality counts associated with a 10.5\% increase in target pollution levels/ 1$^{\circ}$C rise in temperature. The solid red line represents the mean effect estimated from 9000 posterior samples, while the red shaded region denotes the 95\% credible interval. 
    The dashed black line at zero indicates the null effect, where no change in mortality is observed.}
    \label{fig:main}
\end{figure}

Based on the selection procedures, the optimal model specification retained both PM\textsubscript{2.5} and PM\textsubscript{10} for COPD, pneumonia and heart failure. PM\textsubscript{10} was not included in the selected model for the neoplasms outcome measure. Changes in mortality counts and their 95\% credible intervals per 10.5\% pollutant increase or 1$^{\circ}$C temperature rise are shown in Figure \ref{fig:COPD-PM10}-\ref{fig:COPD-Temperature} and Figure S3-S5 (\textbf{Supplementary Material}). 
Table \ref{tab:impact_summary} summarizes the estimated effects of pollution and temperature on the four outcome variables, based on the selection procedures described in Section \ref{Model Selection}. 
For each predictor, the table reports the point estimate and credible interval of the posterior parameter distribution.  Effect estimates from selected models are presented as both numerical and percentage changes in mortality counts. Predictors not included in the final selected model determined via WAIC comparison were indicated with a slash (/). To aid interpretation of the target predictor's effect, we computed the change in the outcome associated with a 1-unit increase in temperature or a 1-unit increase in the pollutant predictor on the log scale, which corresponds to a 10.5\% increase on the original scale. This change was averaged over all 9,000 MCMC posterior samples (from 3 chains, each with 3,000 iterations) generated by our JAGS model. The rationale and derivation of this approach are provided in \textbf{Supplementary Material} Section 4.
\begin{table*}[!t]
\scriptsize
\centering
\renewcommand{\arraystretch}{1.0}
\setlength{\tabcolsep}{1pt}
\caption{Pollutants and Temperature Effects (with 95\% credible interval) Obtained from the Model with Measurement Error}
\begin{tabular}{l c c c c c c c}
\toprule
\textbf{Outcome} & \textbf{PM$_{2.5}$} & \textbf{PM$_{10}$} & \textbf{SO$_{2}$} & \textbf{NO$_{2}$} & \textbf{CO} & \textbf{O$_{3}$} & \textbf{Temperature} \\
\midrule
$C^a$   & 0.01 (-0.86,0.88) & -1.20 (-1.76,-0.62) & / & /  & 0.36 (-0.47,1.22) & -0.45 (-1.44,0.59) & 0.08 (-0.19, 0.37)\\
$C^b$   & 0.03\% (-2.5, 2.23) & -3.03\% (-4.43,-1.58)& / & /  & 0.91\% (-1.17,3.08) & -1.14\% (-3.63,1.48) & 0.21\% (-0.47, 0.93)\\
\midrule
$P^a$   & 0.43 (-0.65,1.52) & -1.37 (-1.90,-0.83) & 0.27 (-0.09,0.62) & 0.51 (-1.97,2.88) & -0.26 (-2.96,2.82) & / & -0.07 (-0.36,0.23) \\
$P^b$   & 1.08\% (-1.66,3.83) & -3.46\% (-4.78,-2.11) & 0.67\% (-0.23,1.56) & 1.27\% (-4.95, 7.26) & -0.65\% (-7.48, 7.13) & / & -0.17\% (-0.90,0.59) \\
\midrule
$H^a$   & -0.48 (-1.22,0.26) & -0.02 (-0.52, 0.45) & -0.37 (-0.61, -0.13) & / & 0.37 (-0.37, 1.13) & -0.75 (-1.60, 0.06) & -0.34 (-0.60, -0.07) \\
$H^b$   & -1.13\% (-2.85,0.60) & -0.05\% (-1.21, 1.05) & -0.86\% (-1.43,-0.30) & / & 0.86\% (-0.87,2.63) & -1.75\% (-3.65,0.14) & -0.79\% (-1.40,-0.16) \\
\midrule
$N^a$   & -0.26 (-0.61,0.08) & / & / & / & / & -0.76 (-1.08,-0.44) & / \\
$N^b$   & -0.59\% (-1.39,0.18) & / & / & / & / & -1.70\% (-2.41,-0.99) & / \\
\bottomrule
\end{tabular}

\medskip
\footnotesize
$^a$ numerical effect of pollutants and temperature ($\#$ of excess deaths/week); $^b$ percentage effect of pollutants and temperature (\%). \\
C: COPD; P: Pneumonia; H: Heart Failure; N: Neoplasms.

\label{tab:impact_summary}
\end{table*}

\begin{table*}[!t]
\scriptsize
\centering
\renewcommand{\arraystretch}{1.0}
\setlength{\tabcolsep}{1pt}
\caption{Pollutants and Temperature Effects (with 95\% credible interval) Obtained from the Model without Measurement Error}
\begin{tabular}{l c c c c c c c}
\toprule
\textbf{Outcome} & \textbf{PM$_{2.5}$} & \textbf{PM$_{10}$} & \textbf{SO$_{2}$} & \textbf{NO$_{2}$} & \textbf{CO} & \textbf{O$_{3}$} & \textbf{Temperature} \\
\midrule
$C^a$   & 0.23 (-0.22,0.68) & -0.61 (-0.92, -0.31) & / & /  & 0.15 (-0.32, 0.63) & -0.30 (-0.75, 0.17) & -0.30 (-0.51, -0.08)\\
$C^b$   
&0.57\% (-0.63,1.73) &-1.52\% (-2.26,-0.77)&/&/&0.34\% (-0.86,1.56)&-0.78\% (-2.00,0.44)&-0.76\% (-1.29,-0.25)\\
\midrule
$P^a$   &0.66 (0.25, 1.08)&-0.62 (-0.88,-0.36)&0.21 (0.05, 0.38)&0.02 (-0.68,0.72)&0.12 (-0.71,0.96)&/&-0.56 (-0.75, -0.36) \\
$P^b$   &2.16\% (0.80,3.55)&-2.00\% (-2.83,-1.17)&0.71\% (0.17,1.26)&0.06\% (-2.18, 2.31)&0.28\% (-2.24,2.94)&/& -1.85\% (-2.47,-1.24)
\\
\midrule
$H^a$   &-0.35 (-0.82, 0.13)&-0.02 (-0.35,0.31)&-0.22 (-0.38,-0.05)&/&0.53 (0,1.06)&-0.35 (-0.83,0.14)&-0.43(-0.67,-0.20) \\
$H^b$   &-0.89\% (-2.00,0.22) &-0.03\% (-0.80,0.74)&-0.51\% (-0.90,-0.12)&/&1.29\% (0.04,2.56)&-0.83\% (-2.03,0.37)& -0.99\% (-1.48,-0.41)\\
\midrule
$N^a$   & -0.13 (-0.38,0.12) & / & / & / & / & -0.66 (-0.95,-0.38) & / \\
$N^b$   & -0.29\% (-0.84,0.26) &/&/&/&/&-1.50\% (-2.12,-0.86)& /\\
\bottomrule
\end{tabular}

\medskip
\footnotesize
$^a$ numerical effect of pollutants and temperature ($\#$ of excess deaths/week); 
$^b$ percentage effect of pollutants and temperature (\%). \\
C: COPD; P: Pneumonia; H: Heart Failure; N: Neoplasms.

\label{tab:impact_summary w.o ME}
\end{table*}

\begin{figure}[!t]
  \centering
  \includegraphics[width=0.9\linewidth]{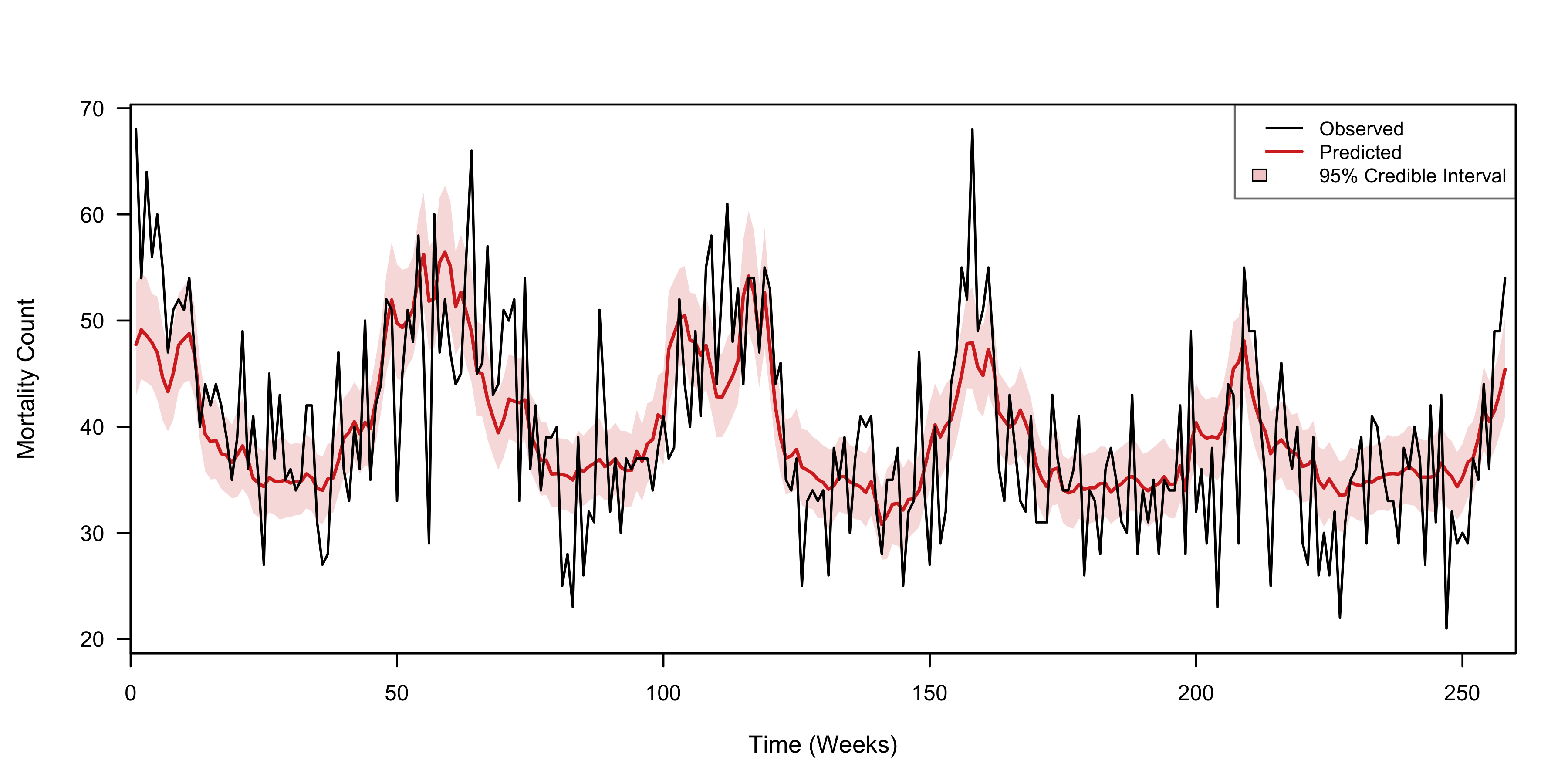}
  \caption{Predicted vs. Observed COPD Mortality using the measurement error model.}
  \label{fig:COPD Mortality}
\end{figure}

Following the model selection strategy outlined in Section~\ref{Model Selection}, we explored the candidate model space using approaches analogous to frequentist stepwise procedures. Specifically, we applied forward selection, backward elimination, and bidirectional selection—adding or removing predictors based on the Widely Available Information Criterion (WAIC) for model comparison. Figures~\ref{fig:COPD-PM10} through~\ref{fig:COPD-Temperature} display the changes in predicted mortality counts when varying each predictor of interest while holding all others constant. 

\begin{itemize}
    \item \textit{COPD}: For COPD mortality, the model selected via backward elimination yielded the lowest WAIC range. The final model included PM\textsubscript{2.5}, PM\textsubscript{10}, CO, O\textsubscript{3} and observed temperature. In terms of percentage change, a 10.5\% increase in PM\textsubscript{2.5} was associated with a -0.03\% change (95\% CI: -2.50\%, 2.23\%) in COPD mortality, holding other variables fixed. Similarly, a 10.5\% increase in PM\textsubscript{10} corresponded to a -3.03\% change (95\% CI: -4.43\%,-1.58\%), a 10.5\% increase in CO to a 0.91\% change (95\% CI: -1.17\%,3.08\%), and a 10.5\% increase in O\textsubscript{3} to a -1.14\% change (95\% CI: -3.63\%,1.48\%). A 1$^\circ$C increase in temperature was associated with a 0.21\% change (95\% CI: -0.47\%, 0.93\%). Figure~\ref{fig:COPD Mortality} compares the averaged predicted mortality counts with the observed COPD mortality over the five-year study period, indicating that the mortality time series was well recovered by the model.

\item \textit{Pneumonia}: For pneumonia mortality, the selected model included all pollutants with the exception of O\textsubscript{3}, and observed temperature. 

\item \textit{Heart failure}: For heart failure, the final model comprised PM\textsubscript{2.5},
PM\textsubscript{10}, SO\textsubscript{2}, 
CO, O\textsubscript{3}, and observed temperature.

\item \textit{Neoplasm}: For neoplasm mortality, the model selection process consistently favoured the inclusion of both PM\textsubscript{2.5} and O\textsubscript{3} as this combination yielded the lowest WAIC range. Furthermore, temperature did not appear to account for the distinctive white-nose pattern observed in neoplasm deaths, suggesting that it plays a negligible explanatory role.

\end{itemize}

For the first three outcome measures, replacing observed temperature with a threshold-adjusted temperature variable resulted in an increased WAIC range, that this model is not preferred.

As explained in Section \ref{simulation},  based on our simulation studies, we found that when the data were generated from the true model (with measurement error) and we fit the correctly specified model (including measurement error), the WAIC was smaller, indicating better model fit as expected. 

For the real data analysis, we applied both model specifications (with and without measurement error) to four different health outcome measures, with the corresponding selected model specification mentioned before in this Section. For all four outcome measures, the measurement error model came out with lower WAIC values than the standard model (Table \ref{tab:WAIC comparison}), which suggests a preferred fit and
superior out-of-sample predictive performance.

\begin{table}[htbp]
\centering \large
\caption{WAIC 
ranges of 10 replicate MCMC analyses 
for four outcome measures under models with and without measurement error structure. 
}
\label{tab:waic_grouped}
\begin{tabular}{l cc}
\toprule
 & \multicolumn{2}{c}{\textbf{WAIC range}} \\
\cmidrule(lr){2-3}
\textbf{Outcome measures} & \textbf{No measurement error} & \textbf{Measurement error} \\
\midrule
COPD & 
(1799.86, 1800.02)
& 
(1760.34 ,  1761.02)\\
Pneumonia & 
(1750.73, 1751.17)
&
(1704.20 ,  1705.16)
\\
Heart Failure & 
(1740.19, 1740.53)
& 
 (1731.89 ,  1732.12)
\\
Neoplasm & 
(1689.19, 1689.37) 
& 
(1688.23 ,  1688.37)\\
\bottomrule
\end{tabular}
\label{tab:WAIC comparison}
\end{table}

The reason why the WAICs of the measurement-error model and non-measurement-error model are comparable are explained in \textbf{Supplementary Materials} Section 3.
Still, the coefficient estimates from the model without measurement error correction (Table \ref{tab:impact_summary w.o ME}) end up being very similar in magnitude to the ones reported in the corrected model (Table \ref{tab:impact_summary}), even though the statistical significance changes in a notable way. In particular, under the no‑measurement‑error assumption, several pollutants show statistically significant positive associations, whereas no such positive effects were seen in the measurement error model. Also, temperature looks statistically significant in a negative direction for some of the outcome measures. Crucially, we will not immediately interpret these negative pollutant coefficients as evidence of a ``protective'' health effect. Rather, they reflect the consequences of fitting a model that treats the observed exposure data as a noisy proxy for the true underlying exposure. Some mismatch between the two sets of results points to how sensitive the inference is to how exposure measurement error is handled.  We discuss further in the next section.

\section{Discussion and Extensions}\label{Discussion}
\subsection{Explaining the Counterintuitive Results}
From Section \ref{Results}, we found that certain pollutants appeared to exert ``protective'' effects on health. While this finding seems counterintuitive, it is not without precedent; researchers have long recognized this phenomenon in the environmental epidemiology literature, where it is commonly referred to as the ``negative coefficient paradox''. This observation further underscores the complexity of the relationship between air pollution and health outcomes, and points to the potential influence of causal inference factors, such as instrumental variables, mediators, unmeasured confounders, and related constructs \citep{wallner2014worldwide}.
Here, we propose several potential explanations for the paradoxical finding:
\begin{enumerate}[label=(\roman*)]
\item \textit{Causal Mediation:}  We consider a possible explanation of the results based on mediation. let there be \( n_t \) time points indexed by \( t = 1, \dots, n_t \) and \( X_{tj} = X_t \) denote the pollution exposure experienced by individual \( j \) at time \( t \); note that all individuals share the same exposure level at a given time point. Let \( Z_{tj} \) represent the mediator on the causal path from air pollution to mortality counts. In general, \( Z_{tj} \) can be interpreted as the level of preventive health behavior for individual \( j \) at time \( t \). 
This coping behavior can be operationalized as a numerical measure, for example, health care expenditure. Supporting this, Liu \citep{liu2021effect} found that a one-unit reduction in the Air Quality Index (AQI) is associated with nearly US\$74 million in annual savings in respiratory-related outpatient expenditure. Alternatively, \( Z \) can simply be measured as expenditure on healthy food. Huang \citep{huang2025polluted} applied an instrumental variable approach and confirmed a causal relationship between PM2.5 and household spending on healthy foods (e.g., fruits, vegetables, and dairy products), showing that such expenditure increases significantly with rising PM2.5 levels. Similarly, Agarwal \citep{agarwal2025eating} provided causal evidence that a 50 \(\mu\)g/m\(^3\) increase in PM2.5 leads to a 7.51\% increase in healthy food expenditure. This mechanism is particularly relevant in the Chinese context, where the cultural principle of ``food as medicine'' is deeply embedded in daily life. This cultural frame makes dietary adjustment feel more salient and intuitive as a response to environmental threats. On this basis, food consumption serves as a mediating step in the causal path from pollution to health, because households naturally turn to more protective foods when air quality becomes challenging or unstable.  Therefore, in environmental epidemiological studies, health care and healthy food expenditures can function as mediators between air pollution and health outcomes since these behavioral changes could be beneficial to health independently of pollution levels.

Unlike exposure, the value of the mediator varies across individuals:
\begin{equation}\label{eq:Z}
Z_{tj} = \gamma_{zx} X_t + c_z + \varepsilon_{tj}^{(z)}, \quad \varepsilon_{tj}^{(z)} \sim N(0, \sigma_{z}^2)
\end{equation}
where $\epsilon_{tj}^{(z)}$
are independent across individuals and time points. Here $\gamma_{zx}>0$
as we would like to demonstrate that higher pollution is associated with higher health care expenditure. 
Let $Y_{tj}$ be the binary outcome for individual $j$ at time point $t$.
Let $Y_{tj}|X_k, Z_{tj} \sim \text{Bernoulli}(p_{tj})$
where the disease probability for individual $j$ at time $t$, $p_{tj}$, follows a logistic model:
\begin{align*}
P(Y_{tj}=1 \mid X_t, Z_{tj}) &= \text{logit} (\beta_{yz} Z_{tj} + \beta_{yx} X_t + \alpha_0) \equiv \Lambda(\alpha_0 + \beta_{yz} Z_{tj} + \beta_{yx} X_t)
\end{align*}
where $\Lambda(u) = 1/(1+e^{-u})$ is the logistic function.
We assume $\beta_{yx} > 0$ (higher exposure increases mortality risk) and $\beta_{yz} < 0$ (higher health care expenditure reduces mortality risk). The causal relationship among the variables is shown in Figure \ref{fig:Paradox_DAG}.

\begin{figure}[h!]
\centering
\caption{Directed acyclic graph (DAG) illustrating the hypothesized causal pathway from air pollution exposure to health outcomes, mediated by the health care/healthy food expenditure. The diagram depicts the direct effect of pollution on health as well as the indirect effect operating through the health care/healthy food expenditure pathway. \\}
\begin{tikzpicture}[scale=1.5]
    \tikzset{rectnode/.style={draw, rectangle, minimum width=2.2cm, minimum height=0.8cm, align=center}}
    
    \node[rectnode] (x) at (-1.5,-1) {Air Pollution};
    \node[rectnode] (z) at (0,1) {Health Care/Healthy Food Expenditure};
    \node[rectnode] (y) at (2.5,-1) {Health Outcomes};

    \path[->](x) edge (z);
    \path[->](z) edge (y);
    \path[->](x) edge[bend right=15] (y);
\end{tikzpicture}
\label{fig:Paradox_DAG}
\end{figure}
\noindent
For the aggregated-level causal effect ($T_{\text{causal}}$) and marginal effect ($T_{\text{marginal}}$) of $X$ on $Y$ under the relationship structure illustrated in Figure \ref{fig:Paradox_DAG}, with the detailed derivation steps in \textbf{Supplementary Material} Section 5, we obtained:
\begin{equation}\label{Causal=Marginal}
\boxed{T_{\text{causal}}=T_{\text{marginal}} \approx \beta_{yx} + \beta_{yz} \gamma_{zx} 
}
\end{equation}
We presented this derivation because we believed that, in practice, mediator variables may be observable at the individual level yet unavailable at the aggregated level. As a result, we had no means of including them when modeling at the aggregate level. This limitation is closely tied to the concept of the ecological fallacy: a logical error in statistical reasoning that occurs when one draws conclusions about individuals solely from aggregate or average data pertaining to the group to which they belong \citep{loney2014individualistic}. Furthermore, this issue illustrated the non-collapsibility \citep{daniel2021making} inherent in the Poisson regression model. Since this model was evaluated on the logarithmic scale, the addition or removal of covariates altered the baseline risk, which in turn shifted the estimated effect of the exposure.

\item  \textit{Marginalization paradox}: Based on Result 
\ref{Causal=Marginal}, the causal effect and the marginal effect are conceptually distinct but numerically identical due to the assumption of no confounders. This helps explain how some variables may act as mediators along the pathway through which air pollution affects mortality counts, yet lead to the marginal effect of air pollution being counterintuitive
(i.e., negative coefficient) 
, as reported in Sections~\ref{Results}. 
This explanation can also be generalized to settings with multiple exposure variables, where each exposure variable is treated as independent from one another and may have either direct or indirect impacts on the outcome.

\item \textit{Biological Explanation (Hormesis):}
Although the majority of the literature argues that air pollution is almost always harmful to human health, even at low exposure levels, this consensus may obscure important non-linearities. Emerging evidence suggests that air pollutants may exhibit paradoxical, non-linear dose-response relationships with health outcomes, challenging the conventional assumption that lower pollution always yields better health. At the biological level, this phenomenon is termed \textit{hormesis}, which describes a positive (beneficial) response of an organism to very low doses of a toxicant, while higher doses induce negative effects   \citep{rillig2025hormesis}. 
As reviewed in the hormesis literature for fine particulate matter (PM$_{2.5}$), the true concentration-response function may instead be J-shaped \citep{cox2012hormesis}.
Hormesis suggests that low concentrations may have no negative impact or could even be beneficial by activating protective biological defences (e.g., antioxidant production) without overwhelming them, while only higher concentrations cause harm. If this J-shaped relationship is correct, further reducing PM$_{2.5}$ at and below current ambient levels could produce no additional health benefits, or could even be beneficial, potentially increasing mortality risks. 

In the air pollution and cancer survival literature, empirical studies have documented such paradoxical patterns. For instance, a study of 1,591 patients in South Korea~  \citep{huh2024air}, an analysis from Pennsylvania~  \citep{mckeon2022air}, and a 2016 study from Los Angeles County, California~  \citep{eckel2016air} all report non-linear dose--response curves. Specifically, U-shaped relationships were observed for O$_3$ and PM$_{10}$ exposures, with thresholds around 20~ppb and 25~$\mu$g/m$^3$, respectively, across different cancer stages (localized, regional, and distant). In Table \ref{tab:mean}, it is evident that in our real data, a great portion of both PM$_{2.5}$ and O$_3$ lies below the corresponding thresholds, indicating relatively low observed pollution levels. Rossnerova \citep{rossnerova2020molecular} also indicated that there is biological evidence that under low pollutant concentrations, cells might turn on mild stress signals, like a preparatory exercise, which then includes up-regulation of antioxidant routes, heat shock proteins, and DNA repair systems, and together this can improve resilience against oxidative harm. But once the exposure levels move past a certain threshold, those same guard mechanisms can get saturated or just become overburdened, and then oxidative stress shows up along with inflammation, plus downstream cellular injury, so the overall outcome looks like higher mortality risk. 



\noindent
\begin{figure}[h!]
\centering
\caption{Directed acyclic graph (DAG) illustrating the hypothesized causal pathways from methyl nitrite and ozone exposure to health outcomes, with methyl nitrite specified as a mediator. Sunlight is modeled as a common cause of both methyl nitrite formation and ozone levels, while ozone and methyl nitrite are each assumed to have direct effects on health.\\}
\begin{tikzpicture}[scale=1.5]
    \tikzset{rectnode/.style={draw, rectangle, minimum width=2cm, minimum height=0.8cm, align=center}}
    
    \node[rectnode] (m) at (-3,0) {Solar Radiation};
    \node[rectnode] (z) at (-1,1) {Methyl Nitrite};
    \node[rectnode] (x) at (-1,-1) {Ozone Level};
    \node[rectnode] (y) at (1,0) {Health Outcome};

    \path[->](m) edge (x);
    \path[->](m) edge (z);
    \path[->](z) edge (y);
    \path[->](x) edge (y);
\end{tikzpicture}
\label{fig:Ozone}
\end{figure}

\item \textit{Confounding by Socioeconomic Factors:} Socioeconomic factors are arguably the most potent source of confounding in this type of ecological analysis.
A simple comparison might suggest that areas with higher levels of air pollution are more industrialized. However, these same areas often benefit from stronger economies, greater wealth, and more advanced healthcare systems; these advantages can significantly reduce overall mortality. This phenomenon, described as the ``prosperity paradox''   \citep{wallner2014worldwide},
creates a critical confound: wealth that is associated with higher pollution exposure is also associated with better health outcomes. This dynamic provides a compelling explanation for the counterintuitive negative association observed in our analysis. For instance, Wallner \citep{wallner2014worldwide}   reported that a model solely considering PM\textsubscript{10} exposure yielded a negative association with cancer mortality; however, after adjusting for economic and healthcare-related indicators, the association became positive. They further noted that smoking prevalence tends to be higher in wealthier countries, adding another layer of behavioral complexity. This pattern, however, is not universal. In regions like Europe or North America, lower socioeconomic status (SES) is typically associated with higher pollution exposure. Conversely, research from rapidly industrializing nations, for example,  China has revealed the opposite: higher SES is linked to greater ambient air pollution exposure, as economic development concentrates both wealth and industrial emissions   \citep{xu2024does}. This was also mentioned in the article \href{https://www.ce.washington.edu/news/article/2022-06-08/higher-socioeconomic-status-linked-increased-air-pollution-exposure-china}{``Higher socioeconomic status linked to increased air pollution exposure in China''}.
This reversal underscores that the direction of confounding is context-dependent, but the principle remains the same.
The relationship is further complicated by its bidirectional nature; evidence shows that environmental pollution can negatively impact residents' income by reducing subjective well-being and labor employment   \citep{ramzan2022environmental}, creating a reinforcing loop that makes causal inference even more challenging.

\item \textit{Confounding in Relation to Ozone}:
At the chemical level, studies examining low ambient ozone and morbidity have sometimes reported negative coefficients, which is termed the Paradoxical Ozone Association (POA): a paradox observed only in regions with methyl ethers in gasoline, with the most plausible explanation being methyl nitrite (MN), a toxic pollutant rapidly destroyed by solar radiation that could confound ozone-health effect models  \citep{joseph2007paradoxical,alves2010air}. The relationship among the variables is shown in Figure \ref{fig:Ozone}. Based on the DAG, solar radiation is a latent confounder because it is an unmeasured common cause of both methyl nitrite and ozone levels, while also ultimately affecting health through both exposure pathways. Since solar radiation is not observed in the analysis, it opens back-door paths from each exposure to the health outcome, meaning that any estimated association between methyl nitrite or ozone and health may be biased by the shared influence of solar radiation rather than reflecting purely causal effects of the exposures themselves.
However, this chemical confounder is no longer relevant to American studies any more
as described in the article \href{https://www.eia.gov/todayinenergy/detail.php?id=36614#}{``The United States continues to export MTBE, mainly to Mexico, Chile, and Venezuela''}: methyl tert-butyl ether (MTBE) was phased out in the late 2000s as a result of water contamination concerns.

\end{enumerate}

\subsection{Future Work}
Given the observational nature of studies on health effects of air pollution, disentangling the effects of one pollutant from others is difficult due to the correlation among the pollution variables   \citep{dominici2003health}, as we mentioned in the Table \ref{tab: Correlation Matrix}.
Interpretation of the effect of an individual pollutant on cause-specific health outcome can be problematic, since multicollinearity can alter parameter estimates   \citep{dominici2003health}. 
The reported estimates are often the effect of, for instance, PM\textsubscript{10}, when all the other pollution levels are held fixed. However, region‑specific analyses have revealed considerable heterogeneity in pollution effects on cause‑specific mortality, with some cities showing different magnitudes or even reversed signs  \citep{samet2000national}.  The pooled (aggregated) estimate may also differ from the city‑specific results, and these patterns vary by cause of death.

In this study, the negative association we observed between some air pollutants and mortality should be taken with substantial caution. It’s tempting to treat it as a true protective effect, but it seems more plausible that it comes from measurement limitations, not from anything actually protective. If these factors are not handled properly, the failure to control them well could give us misleading conclusions about both the direction  and the strength of the pollution—mortality relationship. More broadly, epidemiological analyses using pooled or aggregated data tend to be vulnerable to ecological bias, where what’s seen at the group level doesn’t necessarily match what happens at the individual level, which is the core issue here   \citep{shaddick2013ecological,macdonald2025objectively}. This is also linked with Robinson’s paradox, where correlations built from aggregated sources can be quite different, and even flip in sign, compared with correlations from individual-level analyses   \citep{robinson1952some,robinson2009ecological}. Also, our study period fell during the COVID-19 pandemic; that overlap may have changed both the pollution patterns and the way mortality relates to them. During the pandemic, there were major swings in mobility and healthcare use, and that could have affected the health outcomes we report, including the possibility of excess deaths and disruptions in the mortality reporting systems, which then makes interpretation harder. In particular, some deaths might have been misclassified or logged under broader categories, so cause-specific mortality analyses may face outcome misclassification bias.

Finally, emerging evidence indicates that pollution-health associations are rarely linear. Studies have documented U- and J-shaped concentration-response functions for pollutants such as ozone and PM\textsubscript{2.5}, where health risks do not monotonically increase with exposure but instead exhibit thresholds, inflection points, or even increased risks at low concentrations that defy conventional expectations. These nonlinearities have critical implications.
Future methodological development, such as penalized splines and distributed lag nonlinear models, should therefore continue to explore flexible modeling strategies that can better represent these complex patterns and support more accurate health impact assessments.
\section*{Acknowledgments}
None

\section*{Funding}
This work was supported by a bourse de ma\^{i}trise en recherche (345027) from the Fonds de recherche du Québec (FRQNT), and a Discovery Grant from the Natural Sciences and Engineering Research Council of Canada (NSERC)

\section*{Supplementary Material}
All the simulations and data analysis were implemented in R. The code of this study is accessible through \href{https://github.com/Yanfei1001/Air-Pollution-and-Health}{GitHub}.

\bibliographystyle{plainnat} 
\renewcommand{\refname}{References}
 \bibliography{arXivReferences} 
 
\renewcommand{\thesection}{\Alph{section}}
\end{document}